\documentclass[preprint2]{aastex}

\begin{document}


\title{The Distribution of Bar and Spiral Arm Strengths in Disk Galaxies}


\author{R. Buta and S. Vasylyev}
\affil{Department of Physics and Astronomy, University of Alabama, Box 870324, Tuscaloosa, AL 35487}
\author{H. Salo and E. Laurikainen}
\affil{Division of Astronomy, Department of Physical Sciences, University
of Oulu, Oulu, FIN-90014, Finland}



\begin{abstract}

The distribution of bar strengths in disk galaxies is a fundamental property
of the galaxy population that has only begun to be explored.
We have applied the bar/spiral separation method of Buta, Block, and Knapen
to derive the distribution of maximum relative gravitational bar torques, $Q_b$,
for 147 spiral galaxies in the statistically well-defined
Ohio State University Bright Galaxy Survey (OSUBGS) sample. Our goal is to
examine the properties of bars as independently as possible of their 
associated spirals. We find that the distribution of bar strength declines smoothly
with increasing $Q_b$, with more than 40\% of the sample having $Q_b$ $\leq$ 0.1.
In the context of recurrent bar formation, this suggests that strongly-barred states 
are relatively short-lived compared to weakly-barred or
non-barred states. We do not find compelling evidence for a bimodal
distribution of bar strengths. Instead, the distribution is fairly smooth
in the range 0.0 $\leq$ $Q_b$ $<$ 0.8.

Our analysis also provides a first look at spiral strengths $Q_s$ in the OSU
sample, based on the same torque indicator. We are able to verify
a possible weak correlation between $Q_s$ and $Q_b$, in the sense that galaxies
with the strongest bars tend also to have strong spirals.

\end{abstract}




\section{Introduction}

Bars and spirals are an important part of the morphology
of disk galaxies. These ``showy disk morphological features which
characterize the (Hubble) tuning fork"
(Firmani \& Avila-Reese 2003) play a role in general classification
schemes (e.g., Hubble 1926; Sandage 1961; de Vaucouleurs 1959;
Sandage \& Bedke 1994), and also can be tied to disk galaxy evolution
(e.g., Kormendy \& Kennicutt 2004). Over the past two decades, 
there has been a great
deal of interest in the properties of bars, including quantification
of bar strength (e.g., Elmegreen \& Elmegreen 1985; Martin 1995;
Wozniak et al. 1995; Martinet \& Friedli 1997; Rozas et al. 1998;
Aguerri et al. 1998; Seigar \& James 1998; Aguerri 1999; Chapelon et al. 1999; Abraham \& Merrifield 2000;
Shlosman et al. 2000; 
Buta \& Block 2001; Laurikainen \& Salo 2002; Knapen et al. 2002), bar pattern speeds (Elmegreen et al. 1996; Corsini et al. 2003, 2004; Debattista and Williams 
2004; Aguerri et al. 2003; Debattista et al. 2002; Gerssen et al. 1999; 
Merrifield and Kuijken 1995),
mass inflow rates (Quillen et al. 1995), and studies of
the distribution of bar strengths (Block et al. 2002; Whyte et al.
2002; Buta, Laurikainen,
\& Salo 2004). The most recent studies have indicated, on one hand, that bar
and spiral strength can be quantified in a reasonable manner from
near-infrared images and, on another hand, that such quantifications
are useful for probing both bar and spiral evolution.

The distribution of bar strengths is a particularly important issue.
It is well known that as much as 70\% of normal bright galaxies
are barred at some level (e.g., Eskridge et al. 2002), which
suggests that bars might be long-lived features.
However, in the presence of gas, bars are not expected to be permanent
features of galaxies, but should dissolve in much less than a
Hubble time owing to mass inflow into the nuclear region which
can build up a central mass concentration and destroy a bar (Pfenniger
and Norman 1990). The high frequency of bars has thus led to the
idea that bars dissolve and reform many times during a Hubble time
(Combes 2004).
If this is the case, the distribution of bar strengths will tell us
the relative amount of time a galaxy stays in a given bar state
(strong, weak, or non-barred; Bournaud \& Combes 2002; Block et al.
2002).

Block et al. (2002) and Buta, Laurikainen, \& Salo (2004=BLS04) used the
gravitational torque method (GTM; Buta \& Block 2001; Laurikainen
\& Salo 2002) to derive 
maximum relative nonaxisymmetric torque strengths $Q_g$ for the
Ohio State University Bright Galaxy Survey (OSUBGS, Eskridge et al. 2002), 
a statistically
well-defined sample of nearby bright galaxies. Block et al. (2001, 2004),
BLS04, and Laurikainen, Salo, \& Rautiainen (2002) showed that $Q_g$ 
correlates with deprojected bar ellipticity, a popular parameter
suggested by Athanassoula (1992) to be a useful (although incomplete)
measure of bar strength (e.g., Martin 1995; Whyte et al. 2002). 
The correlation was found by Laurikainen, Salo, \& Rautiainen (2002)
to be much better when objectively-measured near-IR ellipticities are 
used as opposed to the optical ellipticities estimated by Martin (1995)
from blue light photographs. The good correlation is very important, 
because the shape of the bar relates to
the shape of the orbits which build up the bar, which
should depend on the global force field.
Also, BLS04 found that $Q_g$ correlates well
with the bar ellipticity parameter $f_{bar}$ measured by Whyte et al. (2002).
$Q_g$ is a bar strength indicator that is sensitive to the mass of the bar, and
as such should be a better measure of bar strength than
bar ellipticity. However, $Q_g$ is affected
also by spiral arm torques which can dominate over the torques
due to weak bars. Thus, $Q_g$ alone cannot tell us the actual
distribution of bar strengths, but only the distribution for
stronger bars.

One way to derive the distribution of real bar strengths is
to remove the spiral contribution to $Q_g$. Buta, Block, and
Knapen (2003=BBK03) developed a Fourier-based method of separating
bars from spirals that utilizes a symmetry assumption (section 3).
Block et al. (2004) applied this method to deep near-IR images of
17 bright galaxies to derive true bar strengths $Q_b$ and spiral
strengths $Q_s$. This analysis detected a possible correlation 
between $Q_b$ and $Q_s$ in the sense that
among bars having $Q_b$ $>$ 0.3, spiral strength 
increases with increasing bar strength. Block et al. suggested that the
apparent correlation implies that for stronger bars, the bar and the
spiral grow together and have the same pattern speed.

Our goal with the present paper is to apply the BBK03 method to
nearly 150 spiral galaxies in the OSUBGS,
a database of $H$-band (1.65$\mu$m) images that have enough depth
of exposure to allow reliable Fourier analyses. 
In the $H$-band, the extinction is only 19\% that in the $V$-band 
(Cardelli, Clayton, \& Mathis 1989), and such images are suitable for the 
derivation of gravitational potentials using fast-Fourier transform
techniques (Quillen, Frogel, and Gonz\'alez 1994; Salo et al. 1999;
Laurikainen \& Salo 2002). From the separated
images, we derive the distributions of bar and spiral strengths
and investigate what these tell us about disk galaxies.
We also further investigate the correlation between $Q_b$ and
$Q_s$.

\section{Galaxy Sample}

Our sample consists of 147 bright galaxies drawn from the same sample as
used by BLS04, Laurikainen, Salo, and Buta (2004a=LSB04), and 
Laurikainen et al. (2004b=LSBV04).
These previous studies used 180 galaxies, including 158 OSUBGS galaxies 
having total magnitudes $B_T$ $<$ 12.0, $D_{25}$
$<$ 6\rlap{.}$^{\prime}$5, 0 $\leq$ $T$ $\leq$ 9, inclination $<$ 65$^{\circ}$,
and $-$80$^{\circ}$ $<$ $\delta$ $<$ +50$^{\circ}$. In addition, this
sample included 22 galaxies from the Two Micron All-Sky Survey (2MASS,
Skrutskie et al. 1997) which satisfy similar criteria as the OSUBGS
but are larger than the 6\rlap{.}$^{\prime}$5 diameter limit. However,
the 2MASS images are sufficiently underexposed that they proved inadequate
for bar/spiral separation. Whereas bars are detected fairly well in such
images, the spirals and background disks are often too faint to characterize 
reliably, and we do not use these images further in this paper.

\begin{figure}
\figurenum{1}
\plotone{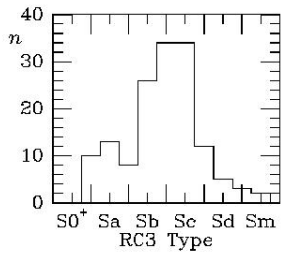}
\caption{Histogram of the distribution of revised Hubble types for the
147 galaxies in our bar/spiral separation sample. The types are from
RC3 (de Vaucouleurs et al. 1991).}
\label{histtypes}
\end{figure}

Figure~\ref{histtypes} shows that our subset of 147 OSUBGS galaxies
is dominated mainly by Sbc and Sc galaxies. The base OSU sample is typical of
the bright galaxy population, as shown by Eskridge et al. (2002) and
Whyte et al. (2002). BLS04 and LSB04 showed that their OSUBGS/2MASS sample is 
biased mainly against inclusion of very late-type, low luminosity barred spirals
and low surface brightness galaxies. Our subset of 147 galaxies has a 
similar bias.

\section{The BBK03 Technique}

The bar/spiral separation method of BBK03 depends on a simple assumption
concerning the behavior of the relative Fourier intensity amplitudes as
a function of radius in a bar: the relative intensities decline past
a maximum in the same or a similar manner as they rose to that maximum.
This is known as the {\it symmetry assumption}. In a complicated bar and
spiral system, only the rising portion of the symmetric curve is seen,
as in BBK03's example of NGC 6951, and the symmetry assumption allows
the extrapolation of the bar into the spiral region. The assumption is
justified from studies of barred galaxies lacking strong spiral structure.
BBK03 used the study of 6 barred galaxies from Ohta et al. (1990)
and the case of NGC 4394 from the OSUBGS to justify the assumption.
The assumption has found further support in studies of
SB0 and SB0/a galaxies from the Near-Infrared S0 Survey (Buta
et al. 2005; see Buta 2004 for a preliminary summary
of these results). In these cases, the bars in the near-infrared
are observed against only bulge and disk components, so that the
bar is the only significant nonaxisymmetric contribution.
Since we cannot know {\it apriori} the form of any particular bar, we judge
the effectiveness of a given bar/spiral separation by examining 
bar plus disk and spiral plus disk intensity maps (see BBK03).
If the bar length is underestimated or overestimated, we can
detect the failure as positive or negative residuals in the
spiral plus disk image.

\section{Application of the BBK03 technique to the OSUBGS Sample}

The application of the BBK03 technique to the OSUBGS sample
required a number of modifications. First,
the method was developed using deep $K_s$ images with pixel sizes
of 0\rlap{.}$^{\prime\prime}$24 (Block et al. 2004). In contrast, 
the OSUBGS $H$-band
images have pixel sizes ranging from 1\rlap{.}$^{\prime\prime}$11
to 1\rlap{.}$^{\prime\prime}$50 and are noisier at large radii
than the $K_s$ images used by Block et al. (2004). These two
factors complicate separation, but the pixel size problem could
be handled effectively by resampling the images into pixels
one-quarter as large using IRAF routine IMLINTRAN.

In our analysis of the OSUBGS sample, we encountered a greater
range of complications in the relative Fourier intensity curves,
such as multiple bars and the effects of decomposition errors
on central isophotes. Thus, it was necessary to adapt the BBK03 method
to deal with the new complications. The effects of decomposition
errors were particularly serious.
For all separations, we used deprojected images based on bulge/disk/bar
decompositions from LSBV04. In each case, the bulge was assumed
to be spherical, but in those cases where the assumption may be
wrong, the decomposition led to symmetric regions of lower
intensity on each side of the center where too much bulge light was 
subtracted. In about a dozen cases, the problem was sufficiently
serious that bar/spiral separation could not be carried out. In
most cases, the problem could be treated in a two-step separation
process, which also proved very effective for cases
with multiple bars or ovals.

\begin{figure}
\figurenum{2}
\plotone{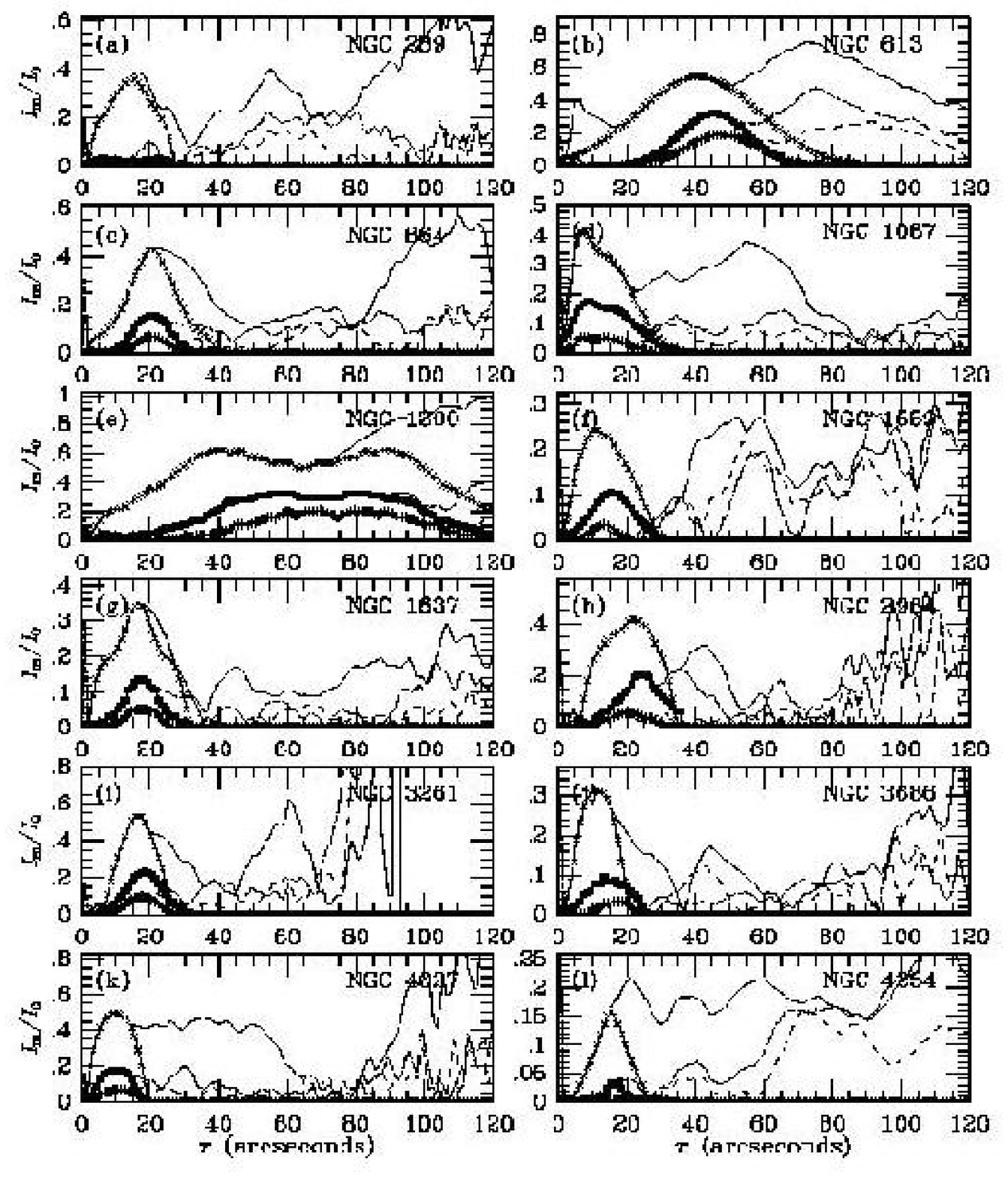}
\caption{}
\label{extraps}
\end{figure}

Figure~\ref{extraps} shows the lower-order Fourier representations used to separate
the bars and spirals in 24 OSUBGS galaxies. These objects illustrate
well the types of extrapolations needed to deal with the wide range
of bar types found in the sample. While for many (e.g., NGC 289, 864, 
1637, 3261, 3686, 4027, 4254, 4303, 4548, 4995, 5085, 5483, 5921, and 6300)
the symmetry assumption appears to work well, for others (e.g., NGC 613, 1087, 4457, 4579,
4593, and the higher order terms for NGC 3261 and 4548)
we found it effective to fit a single gaussian to the rising 
relative intensities (or even a double gaussian, as in NGC 1087).
We also followed the procedure of BBK03 to extrapolate the
bars as little as possible, so that if the observed relative
Fourier amplitudes due to the bar could be detected beyond
the maximum, as much as possible of the decline is used as
observed. Two cases shown in Figure~\ref{extraps} are NGC 1559 and NGC 2964.

It is likely that some bar profiles are indeed gaussian
in nature, although the physical implication of such a representation
is not explored here. Some profiles are symmetric but not necessarily
gaussian (e.g., NGC 1637) or are clearly asymmetric (as in NGC 1087,
1559, and 2964). The effectiveness of the separations of the 24
galaxies are shown in Figure~\ref{separations}. In general, very good
separations are possible by the BBK03 procedure. The partly
gaussian-fitted bar representation for NGC 4548 has cleanly
removed the bar with little residual bar light remaining.
In NGC 4579, gaussian fits to all the main bar Fourier terms
allows the inner part of the spiral to be more clearly seen.
The complex bar in NGC 1087, represented by two non-coincident
gaussians, is well-separated from the complex spiral which itself
appears to be affected by considerable star formation. This bar
does not follow the symmetry assumption, except for the individual
gaussian components.

In many of the bar images, ring-like structures, sometimes slightly 
oval and sometimes weakly spiral, are seen. These rings in large part 
represent the axisymmetric component of the associated spirals. 
Also, some failure of the bar extrapolations can cause these weak
structures. In the spiral images also, one often sees a filling in of
the inner part of the pattern. This is due to the axisymmetric
part of the bar. Maximum relative gravitational torques must
be calculated against the total axisymmetric background,
including whatever contributions the spirals and bars themselves
make to this background.

BBK03 noted that in some bars where the spiral is weak or absent,
the maxima of the higher order bar terms shift towards larger radii 
(e.g., Ohta et al. 1990). These shifts are sometimes seen in our bar 
representations, but in many cases there is little or no shift detectable.
Also, Figure 1 shows that our bar representation for NGC 4593 has
fitted peaks in $m$=4 and 6 at smaller radii than for $m$=2.

The symmetry assumption leads to double-humped bar profiles in the 
strong bars of NGC 1300 and NGC 5921.
In both of these cases, the bar image includes a weak elongated
ring pattern that contributes little to the torques. In both cases also,
the spiral plus disk image looks reasonably bar-free, but a slight
asymmetry in the bar leaves a small residual bar spot on the lower
right end of the bar in NGC 5921.

Separation was especially effective for inner ovals. Small ovals
in NGC 4254, 5085, 5247, 5248, and 5483 were easily mapped and removed
with just a few terms. In the case of NGC 5085, we used the symmetry assumption
twice, one for the inner oval and one for an outer oval (Figure~\ref{extraps}s).
The bar mapping for NGC 5248 shows a double-humped
profile that could in principle be represented as a double-gaussian.
In a few cases (e.g., NGC 613, 4457, 4579, 4593), the separation
successfully removed the primary bar but left a small oval in the center. 
These ovals tend to be weak compared to the primary bar, and were
sometimes left in the spiral plus disk image. In other cases, a
two-step process could be used to remove them from the spiral plus
disk image, if necessary.

\section{Quantitative Bar and Spiral Strengths}

We were able to carry out reasonably successful bar/spiral
separations for 147 OSUBGS galaxies. The 33 missing objects
from the LSBV04 sample include the original 22 2MASS galaxies
in their sample and 11 other cases where the decompositions
left complex residual structure in the bulge region that
prevented a reliable Fourier extrapolation of the bar light. 

For the 147 separated galaxies, we computed bar and spiral strengths
following LSBV04. Gravitational potentials are inferred from
the bar and spiral images assuming a constant mass-to-light
ratio. The potentials are derived from Poisson's equation using 
a fast-Fourier transform technique. A polar grid approach is
used to minimize the effects of noise at large radii (e.g., Laurikainen
and Salo 2002). Vertical thickness is taken into account using
an exponential density function having a scaleheight $h_z$ scaled
from the radial scalelength $h_R$ using a type-dependent formulation
(de Grijs 1998). For each image, the radial variation, $Q_T$, of the maximum 
tangential force relative to the mean background radial force is
computed. Then the maximum ratio from the bar image defines
$Q_b$ and the radius $r(Q_b)$ and the maximum ratio from the
spiral image defines $Q_s$ and $r(Q_s)$. 
Figure~\ref{schematic} shows a schematic of these definitions,
based on NGC 6951 (from BBK03). Since bar/spiral separation
uses mainly even Fourier terms for the bar, the procedure leaves
the odd Fourier terms and much of the image noise in the spiral
plus disk image. Thus, it was necessary to inspect the plots for
the spirals to eliminate spurious maxima due to noise at large
radii.

The BBK03 definition of $Q_g$, shown in Figure~\ref{schematic},
differs slightly from that actually used in
BLS04, LSB04, and LSBV04, where $Q_g$ was taken to be the maximum $Q_T$ in the
bar/oval region when such features were present, and 
the general maximum $Q_T$ for mostly non-barred spirals. In general,
the differences between the formal $Q_g$ defined by BBK03 and that
used in the previously cited papers will not be large,
but were necessitated by the higher noise level in the OSUBGS images
compared to those used by BBK03 and Block et al. (2004).

Table 1 lists the derived parameters for the 147 galaxies. The typical
uncertainties in maximum relative gravitational torques are discussed
by BBK03, BLS04, and LSBV04. In the present paper, we
note that the uncertainty in the constant mass-to-light ratio assumption,
as well as the effects of disk truncation,
will likely affect $Q_b$ and $Q_s$ differently, because $r(Q_s)$ can 
significantly exceed $r(Q_b)$, as shown for many cases in Table 1.
In BLS04, we showed that the typical dark halo correction to $Q_g$
is about 5\% for the galaxies that define the OSUBGS sample, based
on a ``universal rotation curve" analysis (Persic, Salucci, \& Stel
1996). Since $r(Q_g)$ will generally be intermediate between $r(Q_b)$
and $r(Q_s)$, we expect the dark halo contribution affects $Q_s$
more than $Q_b$. To minimize the effects of disk truncation, we chose
a maximum radius (RADMAX) for all calculations of 127 pixels, 
which is the maximum circle contained in each image. RADMAX is the
radius of the zone for which and from which forces are calculated. Laurikainen \&
Salo (2002) showed that as long as RADMAX/$r(Q_g)$ $>$ 2, disk truncation
should not significantly affect the force ratio. In almost all cases,
$r(Q_b)$ satisfies this condition, but $r(Q_s)$ may or may not satisfy it.
Thus, our derived $Q_s$ values cannot be taken
as definitive, but as indicative of the approximate spiral strength. Some
$Q_s$ values are also affected by star formation, and in general $Q_s$
is probably overestimated on average in our analysis.

\section{Bar/Spiral Strength Correlations}

We examine correlations between our measured $Q_b$ and $Q_s$ values
and other parameters. Figure~\ref{byfam} shows plots of $<Q_b>$ and
$<Q_s>$ versus RC3 family classification. Figure~\ref{byfam}a shows
a virtually linear correlation between $<Q_b>$  and RC3 family over
all types, verifying that the intermediate de Vaucouleurs family
class SAB is justified (see also Block et al. 2001). This is the case 
even when the sample
is divided into types $T\leq 4$ (Figure~\ref{byfam}c) and $T> 4$
(Figure~\ref{byfam}e). The right panels of Figure~\ref{byfam} show
only weaker correlations of $<Q_s>$ with bar family. In all three
panels, $<Q_s>$ is higher for SB galaxies than for SA galaxies.
Table 2 summarizes the mean values displayed.

Table 1 also lists the ``OSU $B$'' and ``OSU $H$" family classifications
from Eskridge et al. (2002), based on visual inspection of the 
OSUBGS $B$-band and $H$-band images.
Eskridge et al. noted that many galaxies classified
as non-barred or weakly barred in RC3 appear more strongly barred 
in the near-IR. We have computed a ``$Q_b$ family" to compare 
with these estimates (see Table 3). We define an SA galaxy to be
one which has
$Q_b$ $<$ 0.05, while an SB galaxy is defined to have $Q_b$ $\geq$ 0.25.
Between these extremes we define classes S$\underline{\rm A}$B, SAB, and 
SA$\underline{\rm B}$ using the notation of de Vaucouleurs
(1963), designed to illustrate the continuity aspect of
bar strength. The different $Q_b$ families do not involve
equal steps in $Q_b$: the SB category involves a much broader
range of bar strengths than does SA, and we give a broader range 
to SAB compared to S$\underline{\rm A}$B and SA$\underline{\rm B}$. 
Comparison between the
$Q_b$ family and the OSU $H$ family shows the two often disagree.
Many OSU $H$ SB galaxies end up as $Q_b$ family SAB because
the bars are really not that strong. Table 4 summarizes 6 galaxies
from Eskridge et al. (2000) noted to have changed classification
from SA to SB in going from the $B$ to the $H$ band. However, the
$Q_b$ family indicates the bars are still weak, even in the near-IR.
Some disagreements occur
for cases where the spiral comes off the ends of the bar with a
large pitch angle, an example being NGC 1042. In this case,
there is an oval that we have interpreted
as all or some of the bar for bar/spiral separation.

In other cases, the $Q_b$ family gives an SAB or SA$\underline{\rm B}$
family for cases which are clearly SB in blue light. Some notable
examples are NGC 4643, 5101, and 5701, all early-type spirals.
In these cases, the bars are simply not as strong
as they appear to be because of the presence of a strong
bulge component, which contributes significantly to the axisymmetric 
background.

Table 5 summarizes a general comparison between the $Q_b$ family and
the other sources of bar family classification, including the 
classification of LSBV04 whereby a ``Fourier bar" is defined to be one where 
the phases of the main $m$=2 and $m$=4 Fourier components are maintained
nearly constant in the bar region. The most striking aspect of the
RC3 comparison is the number of objects classified as SA in RC3 which
have a $Q_b$ family of S$\underline{\rm A}$B, meaning some bar or oval
was detected in the near-IR. A similar comparison is found for the OSU $B$
classifications, which is not surprising since RC3 families are also
based on $B$-band images. The comparison with the OSU $H$ classifications
shows that, as highlighted before, many OSU $H$ SB galaxies have bars
that are not that strong, and come out with a $Q_b$ family of SAB. The
Fourier bar comparison gives very similar results but requires mainly that 
the low order Fourier phases be constant, not that the bar should 
be especially strong.

Figure~\ref{byparams} shows plots of $<Q_b>$ and $<Q_s>$ versus:
(a,b) the extinction- and tilt-corrected absolute blue magnitude $M_B^o$
based on parameters from RC3 and distances from Tully (1988);
(c,d) the extinction and tilt-corrected mean effective blue light 
surface brightness $\mu_{eo}^{\prime}$ (mag arcmin$^{-2}$; 
see equation 71 of RC3); and (e,f) de Vaucouleurs revised Hubble-Sandage 
type, coded on the RC3 numerical scale. Little correlation with
absolute magnitude is found, although this is partly due to the
fact that the sample is biased against late-type, low luminosity
barred spirals (BLS04). Except for the lowest surface brightness
bin, there is little correlation between $<Q_b>$ and $\mu_{eo}^{\prime}$,
while some correlation between $<Q_s>$ and $\mu_{eo}^{\prime}$
is found. The mean spiral strength appears to increase with
increasingly fainter surface brightness, changing by more than a
factor of two, an effect that may be partly due to star formation and
partly due to increased image noise for the lower surface brightness
objects. Figure~\ref{byparams}e shows the same kind of
correlation between $<Q_b>$ and type discussed by BLS04 and LSB04 for $<Q_g>$,
in the sense that maximum relative gravitational torques are larger for later 
types. Figure~\ref{byparams}f shows the same may be true for spiral strengths
as well. 

BLS04 and LSB04 attributed the type dependence in $Q_g$ 
as due to the increased prominence of the bulge in early-type spirals.
This rather counterintuitive result, that significant-looking bars in
early-types actually have weaker relative torques than those in
later types, is due to the fact that
what the eye recognizes in photographs as a ``bar strength"' is the LOCAL SURFACE
DENSITY contrast, which is different from the tangential to radial force
ratio or its maximum value $Q_g$. The latter is a global quantity, measuring 
FORCES from all parts of the galaxy, and should be a more reliable
indicator of actual bar strength. This
highlights the advantage of the GTM in quantifying bar strength
beyond the visual appearance of bars (LSB04). Early-type bars may in fact
be more massive and intrinsically stronger than those in later types,
but relative to the axisymmetric disks they are imbedded within, late-type
bars can be stronger.

Figure~\ref{bymoreparams} shows correlations between $<Q_b>$
and $<Q_s>$ and $logR_{25}$, the RC3 logarithmic standard isophotal
axis ratio (used as an indicator of inclination), and visually estimated
Arm Class (Elmegreen and Elmegreen 1987). 
Arm Classes (ACs) emphasize the symmetry and continuity of spiral arms,
and it is worth investigating whether these might correlate with
$Q_s$. No significant trend of either nonaxisymmetric strength parameter 
is found with $logR_{25}$, confirming that there is no systematic bias 
introduced to the torques due to deprojection corrections.
However, we detect a weak correlation between $<Q_s>$ and AC in the sense
that $<Q_s>$ is higher for ACs of 9 and 12 (there are no ACs of types 10
and 11), which include the most
symmetric, longest arms, compared to ACs of 1--3, which include
the chaotic, fragmented arms seen in flocculent spirals. In spite 
of the apparent correlation, $Q_s$ is not necessarily a suitable
replacement for the AC because there is considerable overlap among the
classes and the two parameters measure different aspects of spiral structure.

Finally, Figures~\ref{bymoreparams}e,f show $<Q_b>$ and $<Q_s>$ for
SB galaxies as
functions of RC3 variety classification: ringed (r), pseuodringed (rs),
and spiral-shaped (s). With our direct estimates of bar strength, we
can investigate the claim made by Kormendy \& Kennicutt (2004) that
SB(r) galaxies have stronger bars than SB(s) galaxies, based on
the hydrodynamic simulations of Sanders \& Tubbs (1980). Table 2
summarizes $<Q_b>$ and $<Q_s>$ for the three varieties separated
by family. Figure~\ref{bymoreparams}e shows that, on average,  
SB(r) galaxies have weaker bars than SB(s) galaxies, contrary
to the conclusion of Kormendy \& Kennicutt. In our sample also,
SB(rs) galaxies have bars as strong on average as those in SB
galaxies. The differences are not that significant owing to the large
scatter at each variety. Table 2 shows that the statistics
are more uncertain for SA galaxies, which strongly emphasize
the (s)-variety, and SAB galaxies, which strongly emphasize the (rs) variety.

\section{Distribution of Bar and Spiral Arm Strengths}

Figures~\ref{histos}a and b show both differential and cumulative
distributions of $Q_b$,  while Figures~\ref{histos}c and d show the
same for $Q_s$,  for 147 OSUBGS galaxies. For
comparison, the distributions of $Q_g$ values (from LSBV04)
for the same galaxies are shown in Figures~\ref{histos}e and f. 
Figures~\ref{histos}a and b show that when bars 
are isolated from spirals in galaxy images, the lowest bar strength
bins, 0.0-0.05 and 0.05-0.10, fill up considerably over the $Q_g$
bins. More than 40\% of the galaxies have bars with $Q_b$ $\leq$ 
0.10, while only 22\% have $Q_g$ $\leq$ 0.10. It is clear that weak bars or 
ovals are often
masked by spirals and not detected via the $Q_g$ parameter; these bars
are visible in $Q_T$ profiles but have force maxima much lower
than those induced by spiral arms in the outer parts of the disks.
Thus, $Q_g$ does not
give a reliable indication of the relative frequency of weak bars.

For the spirals, the lowest $Q_s$ bin is deficient in galaxies compared to the
next highest $Q_s$ bin, which is not
unexpected given that the sample excludes S0 galaxies. Most spirals
are nevertheless fairly weak, with more than 75\% having $Q_s \leq $ 0.20.

These parameters allow us to assign all the sample galaxies to bar and
spiral strength classes (see Table 1). We follow Buta \& Block (2001) to
make these assignments. For bar class 0 we include any galaxy having
$Q_b$ $<$ 0.05, while for spiral class 0 we include any galaxy having
$Q_s$ $<$ 0.05. 
For bar class 1 we include galaxies having 0.05$\leq$ $Q_b$ $<$0.15,
while for spiral class 1 we include galaxies having 0.05$\leq$ $Q_s$ $<$0.15,
etc. Thus, the 0 class for bars and spirals involves a narrower range
since $Q_b$ and $Q_s$ cannot be negative as defined. These spiral and
bar classes define a quantitative near-infrared classification of bars
and spirals and can be incorporated into the dust-penetrated classification
scheme of Block \& Puerari (1999; see Buta \& Block 2001). While bar class
may represent a suitable replacement for de Vaucouleurs family classifications,
spiral class only distinguishes early and late-type spirals and does not
discriminate well between individual $T$-types (Figure~\ref{byparams}f).

\section{Correlation Between $Q_s$ and $Q_b$}

Elmegreen \& Elmegreen (1985) used bar/interbar and arm/interarm
contrasts to show that strong spirals are associated with
strong bars. Although the bulk of their correlation is
based on only a few galaxies, the implication was that the bars might
be driving the spirals. However, Sellwood \& Sparke (1988) used numerical
simulations to show that bars and spirals might be independent features,
with different pattern speeds. Block et al. (2004) applied the BBK03
technique to 17 intermediate to late-type spirals and found some
correlation between bar and spiral arm torques, but only for the 
strongest bars. These authors suggested that in strongly-barred
galaxies, the bar and the spiral may be growing together and have
the same pattern speed.

Figure~\ref{qsqb} shows the correlation between $Q_s$ and $Q_b$ (crosses)
for the OSUBGS sample. For comparison, the values from Block
et al. (2004) for 17 bright spirals are plotted also. The
solid curve shows the medians in $Q_s$ for successive
bins of 0.1 in $Q_b$. The plot shows that the median $Q_s$
increases from 0.1 to 0.30 as $Q_b$ increases from 0.05 to 0.75.
The rise agrees with that found by Block et al. (2004) within
the uncertainties, and again suggests that at lower bar
strengths, spiral and bar strengths are largely uncorrelated,
while at stronger bar strengths some correlation may be present.
The result is difficult to interpret because the numbers of
galaxies decrease significantly with increasing $Q_b$. Also,
$Q_b$ and $Q_s$ have correlated uncertainties, in the sense that
if $Q_b$ is overestimated by the separation procedure, then $Q_s$
will be underestimated and vice versa. These uncertainties could
be reduced with better quality images as used by Block et al. (2004).
Our results largely support
the idea of Sellwood \& Sparke (1988) that spirals and bars are
independent features, with likely different pattern speeds, at
least for $Q_b <$ 0.3. This is not definitive, however, because
as noted by Block et al. (2004), the frequent alignment of bars 
and rings, which are often parts of the spiral pattern, implies
similar pattern speeds in some cases. For higher bar strengths, some correlation
between $Q_b$ and $Q_s$ may be present that can only be confirmed 
with a larger sample of strongly barred spirals.

\section{Discussion}

\subsection{What Determines Bar Strength?}

Bar strength in isolated disk galaxies is thought to be determined largely by the effectiveness
with which a bar can transfer angular momentum to other galactic 
components, such as spiral structure, resonances, live halos, and outer 
bulge stars (Athanassoula
2003). A bar can get very strong if there is nothing to negate this
effect. However, a bar can affect its own evolution by driving gas 
into the center. This builds up the central mass concentration and
can lead to an inner Lindblad resonance (ILR), which will feed angular momentum
to the bar. When this happens, the bar's orbital structure can be destroyed
and the bar itself fades away (Norman, Sellwood, \& Hasan 1996).

Bar strength in non-isolated galaxies can be affected by tidal
interactions (Noguchi 1996; Miwa \& Noguchi 1998 and also
accretion of gas-rich dwarfs or infalling external gas
(Sellwood \& Moore 1999; Bournaud \& Combes 2002; Combes 2004). 
Miwa and Noguchi (1998) have argued that which bar-forming mechanism (spontaneous
or tidal) dominates depends on the relative importance of the
disk and halo. They suggest that spontaneous bars will be important if disks
are massive relative to their halo, while tidally-induced bars will dominate
if the disks are stable against spontaneous bar formation. Noguchi (1987)
suggested that the ``exponential" and ``flat" bars of Elmegreen \& Elmegreen
(1985) are distinguished by these same two mechanisms, with the former being
spontaneous and the latter being tidally triggered. If gas
flow helps to dissolve a bar, an interaction may regenerate
a bar if there is little disk gas remaining (Berentzen et al. 2004).
If a galaxy accretes substantial external gas, it may be susceptible
to multiple or recurring bar episodes (Bournaud \& Combes 2002, 2004). 
Several simulation studies (Athanassoula
2003; Athanassoula \& Misiriotis 2002; Miwa \& Noguchi 1998; Berentzen 
et al. 2004) have found a correlation between bar strength and bar
pattern speed, in the sense that stronger bars have lower pattern
speeds.

These results suggest that bar strength is not a permanent feature of
galaxies, but can be highly variable over a Hubble time. Evidence in support
of this idea comes from the inverse correlation between central mass concentration
and bar ellipticity in a sample of spiral galaxies (Das et al. 2003). Thus, the distribution of
bar strengths in galaxies may be influenced by a complex variety of 
effects: environment, mass distribution, the interstellar medium,
and the properties of dark matter halos.

\subsection{The Distribution of Bar Strengths: Observations versus Theory}

We have shown in this paper that
a straightforward Fourier technique can be used to separate bars 
from spirals, allowing us to examine the distribution of
bar strengths in galaxies unaffected by the torques due to
spirals. We find a preponderance of low bar strengths that was
masked in previous $Q_g$ studies partly because of the effects
of spiral arm torques. $Q_g$ is only reliable as a bar strength
indicator if the bar is the dominant nonaxisymmetric feature in
a galactic disk. In cases where the spiral dominates, or where the
bar and spiral have comparable strengths, $Q_g$ will be an overestimate
of bar strength.

The reason for wanting to look at the distribution of bar strengths
alone is Sellwood's (2000) assertion that ``most real bars are not made
by the bar instability." This global dynamical instability was first
identified in $n$-body models that showed that a disk-shaped galaxy having 
sufficient kinetic energy in ordered rotational motion would be unstable
to the formation of a bar (Sellwood 1996 and references therein). The
way to avoid the linear instability would be a high central concentration,
guaranteeing the existence of an ILR inside the bar. Sellwood
(2000) noted that many strong bars, such as those found in galaxies like
NGC 1300 and 1433, include small circumnuclear rings whose presence has
been tied to the existence of an ILR region (although the exact locations
of the rings may not be coincident with ILR; Regan and Teuben 2003).
Sellwood noted that enough barred galaxies showed these features to
cast considerable doubt on the bar instability as being the explanation
of most bars. Other features of the strong bars that suggest the influence
of ILRs are the shapes of offset dust lanes (Athanassoula 1992) and
observed gas velocity fields. 

Sellwood also brought attention to
the results of early high redshift studies (e.g., Abraham et al. 1999)
that indicated bars are less frequent for $z$ $>$ 0.5, suggesting that 
bars develop long after the disk formed. However, this conclusion has been
refuted by more recent studies (Sheth et al. 2003; Elmegreen, Elmegreen, 
\& Hirst 2004; Jogee et al. 2004), which indicate no significant drop in the 
bar fraction out to $z$$\approx$1. Jogee et al. (2004) present the most 
comprehensive study of bar fraction as a function of redshift, and found
that this fraction is virtually constant at 30\%$\pm$6\% to $z$=1.
The implications of their result are thought to be that cold, unstable
disks were already in place by $z$=1, and that bars must survive
at least 2 Gyr.

In looking for an alternative to the bar instability, Sellwood (2000)
suggested bar growth occurs through an episodic process, where the
interaction between a bar and a spiral can add particles to the
bar and make it longer, while at the same time reducing the bar's 
pattern speed. He suggested that it would be useful to be able to
predict the distribution of bar strengths for various bar formation
scenarios. Of course, it is also useful to know the observed distribution
of bar strengths. The BBK03 method and the OSUBGS have allowed us to
consider this for the first time.

The only theoretical predictions of an expected distribution of bar strengths
have been made for recurrent bar formation models (Bournaud \& Combes
2002). Block et al. (2002) used a preliminary $Q_g$ analysis of the OSUBGS
sample to derive an observed distribution, and then used the Bournaud
\& Combes simulation database to derive a theoretical distribution using
as much as possible the same assumptions: constant mass-to-light ratio;
exponential vertical density distribution having $h_z=1/12 h_R$; inclusion
of spiral torques; bulges assumed as flat as the disks; and
dark matter ignored. These authors noted that the observed $Q_g$ distribution
shows a deficiency of low $Q_g$ galaxies and an extended ``tail" of high
$Q_g$ galaxies. The comparison showed that both characteristics were best
explained if galaxies are open systems, accreting enough external gas to
double their mass in a Hubble time. The distribution of bar strengths would
then mainly tell us the relative amount of time galaxies spend in different
bar states (strong, weak, or non-barred). The deficiency of low $Q_g$ galaxies
was interpreted as due to the ``duty cycle" between bar episodes. That is,
accretion prevented most galaxies from spending much time in a perfectly
axisymmetric state. Some of the nonaxisymmetric torques could be due to spirals
which would also be maintained by accretion.

The refined GTM analysis carried out by BLS04, LSB04, and LSBV04 provided
a more reliable distribution of $Q_g$. BLS04 showed that even with refinements
that account properly for bulge shapes, and using improved estimates of $h_z$
that allow for the type dependence of $h_z/h_R$ as well as values of $h_R$ derived
from two-dimensional bar/bulge/disk decompositions, the observed distribution
of $Q_g$ still shows a deficiency of objects having $Q_g$ $<$ 0.05 and an
extended tail of high $Q_g$ objects. However, the refined distribution shows
more low $Q_g$ values than did the Block et al. (2002) analysis, due 
to a variety of effects discussed by BLS04.

We find that when spiral torques are removed, the distribution of bar strengths is
a relatively smoothly declining function with increasing $Q_b$. It appears that 
galaxies spend more time in a relatively weakly-barred or non-barred state than
they do in a strongly-barred state. Even in these weakly-barred states, they can
have significant spiral torques. The question now is whether gas accretion models
can account for the actual distribution of bar strengths rather than simply the
distribution of total nonaxisymmetry strengths. In principle, a separation
analysis could be made for simulations as for images. 

Whyte et al. (2002) analyzed the blue and near-infrared images in the OSUBGS, and
derived a quantitative bar strength parameter, $f_{bar}$, which is a rescaled measure
of bar ellipticity (Abraham \& Merrifield 2000). For the large and well-defined 
OSUBGS sample, they derived a distribution of $f_{bar}$ which they claim
shows evidence for bimodality, and argue that the bimodality is likely due 
to rapid evolution from the SB phase to SA and SAB phases. However, the 
distribution of $Q_b$ suggests a continuous distribution of bar strengths,
with no evidence of bimodality. The two results are not really in disagreement
because the evidence for bimodality in $f_{bar}$ is very weak, especially
in the plot of $f_{bar}$ versus concentration shown by Whyte et al. (2002).
The original evidence was found in this same kind of plot by Abraham \& Merrifield
(2000). In agreement with our results, Whyte et al. (2002) found that SAB
galaxies have values of $f_{bar}$ intermediate between SA and SB galaxies.

\section{Conclusions}

Using a simple Fourier technique, we have separated the bars and spirals
in 147 OSUBGS galaxies, and for the first time derived the distribution of 
actual bar strengths in disk galaxies. We find that the relative frequency
of bars is a declining function of bar strength, with more than 40\% of the
sample being very weakly-barred or non-barred with $Q_b$ $<$ 0.1. The
higher frequency of weak bars compared to strong ones suggests that
strong bars are either very transient or may require more special conditions,
such as an interaction. If, in fact, bars are long-lived, as suggested by
the results of high redshift studies (e.g., Jogee et al. 2004), then the observed
distribution of bar strengths is telling us that cold, unstable disks
preferentially form weak bars.

An important piece of the whole picture of barred galaxies is still missing:
SB0 galaxies. Block et al. (2002) suggested
that in the absence of gas, bars are very robust and can last a Hubble time.
What is the distribution of bar strengths in such galaxies? Our SB0 survey
(Buta et al. 2005; Buta 2004) should be able to answer this question.

We thank an anonymous referee for helpful comments on the manuscript.
RB and SV acknowledge the support of NSF Grant AST-0205143 to the University of
Alabama. EL and HS acknowledge the support of the Academy of Finland, and
EL also from the Magnus Ehrnrooth Foundation. SV acknowledges the support
of the Academy of Finland during two summer visits to Oulu in 2002 and 2003.
Funding for the OSU Bright Galaxy
Survey was provided by grants from the National Science
Foundation (grants AST-9217716 and AST-9617006), with
additional funding from the Ohio State University.

\clearpage

\centerline{REFERENCES}

\noindent
Abraham, R. G. \& Merrifield, M. R. 2000, \aj, 120, 2835

\noindent
Abraham, R. G., Merrifield, M. R., Ellis, R. S., Tanvir, N. R., \&
Brinchmann, J. 1999, \mnras, 308, 569

\noindent
Aguerri, J. A. L. 1999, \aap, 351, 43

\noindent
Aguerri, J. A. L., Beckman, J. E., \& Prieto, M. 1998, \aj, 116, 2136 

\noindent
Aguerri, J. A. L., Debattista, V., and Corsini, E. M. 2003, MNRAS, 338, 465

\noindent
Athanassoula, E. 1992, \mnras, 259, 328

\noindent
Athanassoula, E. 2003, \mnras, 341, 1179

\noindent
Athanassoula, E. \& Misiriotis, A. 2002, \mnras, 330, 35

\noindent
Berentzen, I., Athanassoula, E., Heller, C. H., \&  Fricke, K.J. 2004, \mnras, 347, 220

\noindent
Block, D. L. \& Puerari, I. 1999, \aap, 342, 627

\noindent
Block, D. L., Puerari, I., Knapen, J. H., Elmegreen, B. G., Buta, R.,
Stedman, S., \& Elmegreen, D. M. 2001, \aap, 375, 761

\noindent
Block, D. L., Bournaud, F., Combes, F., Puerari, I., \& Buta, R. 2002,
\aap, 394, L35

\noindent
Block, D. L., Buta, R., Knapen, J. H., Elmegreen, D. M., Elmegreen, B. G.,
\& Puerari, I. 2004, \aj, 128, 183

\noindent
Bournaud, F. \& Combes, F., 2002 \aap, 392, 83

\noindent
Bournaud, F. \& Combes, F., 2004, in SF2A-2004, Semaine de l'Astrophysique,
F. Combes, D. Barret, T. Contini, F. Meynadier, \& L. Pagani, eds., Ed-P
Sciences, Conference series

\noindent
Buta, R. 2004, in Penetrating Bars Through Masks of Cosmic Dust,
The Hubble Tuning Fork Strikes a New Note, D. L. Block, I. Puerari,
K. C. Freeman, R. Groess, \& E. K. Block, eds., Springer, Dordrecht,
101

\noindent
Buta, R. \& Block, D. L. 2001, \apj, 550, 243 

\noindent
Buta, R., Block, D. L., \& Knapen, J. H. 2003, \aj, 126, 1148 (BBK03)

\noindent
Buta, R., Laurikainen, E., \& Salo, H. 2004, \aj, 127, 279 (BLS04)

\noindent
Buta, R., Laurikainen, E., Salo, H., Block, D. L., \& Knapen, J. H. 2005, in preparation

\noindent
Cardelli, J. A., Clayton, G. C., \& Mathis, J. S. 1989, \apj, 345, 245

\noindent
Chapelon, S., Contini, T., \& Davoust, E. 1999, \aap, 345, 81

\noindent
Combes, F. 2004, in Penetrating Bars Through Masks of Cosmic Dust,
The Hubble Tuning Fork Strikes a New Note, D. L. Block, I. Puerari,
K. C. Freeman, R. Groess, \& E. K. Block, eds., Springer, Dordrecht,
57

\noindent
Corsini, E. M., et al. 2004, IAU Symp. 220, S. Ryder, D. J. Pisano,
M. A. Walker, \& K. C. Freeman, eds., p. 271

\noindent
Corsini, E. M., Debattista, V., \& Aguerri, J. A. L. 2003, ApJ, 599, 29

\noindent
Das, M, et al. 2003, ApJ, 582, 190

\noindent
Debattista, V., Corsini, E. M., and Aguerri, J. A. L. 2002, MNRAS, 332, 65

\noindent
Debattista, V. and Williams, T. 2004, ApJ, 605, 714

\noindent
de Grijs, R. 1998, \mnras, 299, 595

\noindent
de Vaucouleurs, G. 1959, Handbuch der Physik, 53, 275

\noindent
de Vaucouleurs, G. 1963, \apjs, 8, 31

\noindent de Vaucouleurs, G. et al. 1991, Third Reference Catalog
of Bright Galaxies (New York: Springer) (RC3)

\noindent
Elmegreen, B. G. \& Elmegreen, D. M. 1985, \apj, 288, 438

\noindent
Elmegreen, B. G., Elmegreen, D. M., \& Hirst, A. C. 2004, \apj, 612, 191

\noindent
Elmegreen, B. G., Elmegreen, D. M., Chromey, F. R., Hasselbacher, D. A., \&
Bissell, B. A. 1996, \aj, 111, 2233

\noindent
Elmegreen, D. M. \& Elmegreen, B. G. 1987, \apj, 314, 3

\noindent
Eskridge, P., Frogel, J. A., Pogge, R. W., et al. 2000, \aj, 119, 536

\noindent
Eskridge, P., Frogel, J. A., Pogge, R. W., et al. 2002, \apjs, 143, 73

\noindent
Firmani, C. \& Avila-Reese, V. 2003, RMxAC, 17, 107

\noindent
Gerssen, J., Kuijken, K., and Merrifield, M. R. 1999, MNRAS, 306, 926

\noindent
Hubble, E. 1926, \apj, 64, 321

\noindent
Jogee, S. et al. 2004, \apj, 615, L105 

\noindent
Knapen, J., P\'erez-Ram\'irez, D. \& Laine, S. 2002, \mnras, 337, 808

\noindent
Kormendy, J. and Kennicutt, R. 2004, ARAA, 42, 603

\noindent
Laurikainen, E., Salo, H., \& Buta, R. 2004a, \apj, 607, 103 (LSB04)

\noindent
Laurikainen, E., Salo, H., \& Rautiainen, P. 2002, \mnras, 331, 880

\noindent
Laurikainen, E. \& Salo, H. 2002, \mnras, 337, 1118

\noindent
Laurikainen, E., Salo, H., Buta, R., \& Vasylyev, S. 2004b, \mnras, 355, 1251 
(LSBV04)

\noindent
Martin, P. 1995, \aj, 109, 2428

\noindent
Martinet, L. \& Friedli, D. 1997, \aap, 323, 363

\noindent
Merrifield, M. R. \& Kuijken, K. 1995, MNRAS, 274, 933

\noindent
Miwa, T. and Noguchi, M. 1998, ApJ, 499, 149

\noindent
Noguchi, M. 1987, \mnras, 228, 635

\noindent
Noguchi, M. 1996, in Barred Galaxies, IAU Coll. No. 157, R. Buta, D. Crocker,
and B. G. Elmegreen, eds., ASP Conf. Ser. 91, p. 339

\noindent
Norman, C., Sellwood, J. A., \& Hasan, H. 1996, \apj, 462, 114

\noindent
Ohta, K., Hamabe, M., \& Wakamatsu, K. 1990, \apj, 357, 71

\noindent
Persic, M., Salucci, P., \& Stel, F. 1996, \mnras, 281, 27

\noindent
Pfenniger, D. \& Norman, C. A. 1990, \apj, 363, 391

\noindent
Quillen, A. C., Frogel, J. A., \& Gonz\'alez, R. A. 1994, \apj, 437, 162

\noindent
Quillen, A. C., Frogel, J. A., Kenney, J. D., Pogge, R. W., \& Depoy, D. L., 1995, \apj, 441, 549

\noindent
Regan, M. \& Elmegreen, D. M. 1997, \aj, 114, 965

\noindent
Regan, M. \& Teuben, P. 2003, \apj, 582, 723

\noindent
Rozas, M., Knapen, J. H., \& Beckman, J. E. 1998, \mnras, 301, 631

\noindent
Salo, H., Rautiainen, P., Buta, R., Purcell, G. B., Cobb, M. L., Crock
er, D. A., \& Laurikainen, E. 1999, AJ, 117, 792

\noindent
Sandage, A. R. 1961, The Hubble Atlas of Galaxies, Carnegie Inst. of
Wash. Publ. No. 618

\noindent
Sandage, A. \& Bedke, J. S. 1994, The Carnegie Atlas of Galaxies, Carnegie
Inst. of Wash. Publ. No. 638

\noindent
Sanders, R. H. \& Tubbs, A. D. 1980, \apj, 235, 803

\noindent
Seigar, M. S. \& James, P. S. 1998, \mnras, 299, 672

\noindent
Sellwood, J. 1996, in Barred Galaxies, IAU Coll. No. 157, R. Buta, D. Crocker,
and B. G. Elmegreen, eds., ASP Conf. Ser. 91, p. 259

\noindent
Sellwood, J. A. 2000, in Dynamics of Galaxies: From the Early Universe
to the Present, F. Combes, G. A. Mamon, \& V. Charmandaris, eds.,
San Francisco, ASP Conf. Ser. 197, p. 3.

\noindent
Sellwood, J. A. \& Moore, E. M. 1999, \apj, 510, 125

\noindent
Sellwood, J. A. \& Sparke, L. S. 1988, \mnras, 231, 25P

\noindent
Sheth, K., Regan, M. W., Scoville, N. Z., \& Strubbe, L. E. 2003, \apj, 592, 13

\noindent
Shlosman, I., Peletier, R. F., \& Knapen, J. H. 2000, \apj, 535, L83

\noindent
Skrutskie, M. F. et al. 1997, in The Impact of Large-Scale Near-IR Surveys, F. Grazon et al., eds., Dordrecht, Kluwer, p.25

\noindent
Tully, R. B. 1988, Nearby Galaxies Catalogue, Cambridge, Cambridge
University Press

\noindent
Whyte, L., Abraham, R. G., Merrifield, M. R., Eskridge, P. B., Frogel,
J. A., \& Pogge, R. W. 2002, \mnras, 336, 1281

\noindent
Wozniak, H., Friedli, D., Martinet, L., Martin, P., \& Bratschi, P. 1995, \aaps, 111, 115

\clearpage

\begin{deluxetable}{llllccrrccl}
\tabletypesize{\scriptsize}
\tablewidth{0pc}
\tablecaption{Summary of Parameters}
\tablehead{
\colhead{Galaxy} &
\colhead{RC3} &
\colhead{OSU $B$} &
\colhead{OSU $H$} &
\colhead{$Q_b$} &
\colhead{$Q_s$} &
\colhead{$r(Q_b)$} &
\colhead{$r(Q_s)$} &
\colhead{bar} &
\colhead{spiral} &
\colhead{$Q_b$} 
\\
\colhead{} &
\colhead{family} &
\colhead{family} &
\colhead{family} &
\colhead{} &
\colhead{} &
\colhead{(arcsec)} &
\colhead{(arcsec)} &
\colhead{class} &
\colhead{class} &
\colhead{family} 
\\
\colhead{1} &
\colhead{2} &
\colhead{3} &
\colhead{4} &
\colhead{5} &
\colhead{6} &
\colhead{7} &
\colhead{8} &
\colhead{9} &
\colhead{10} &
\colhead{11} 
} 
\startdata
NGC   150       & SB & SAB & SB &     0.475 &     0.254 &    23 &    33 &   5 &   3 & SB                      \\
NGC   157       & SAB & SB & SA &     0.024 &     0.323 &     3 &    21 &   0 &   3 & SA                      \\
NGC   210       & SAB & SA & SB &     0.052 &     0.037 &    29 &    63 &   1 &   0 & S$\underline{\rm A}$B   \\
NGC   278       & SAB & SA & SA &     0.046 &     0.064 &     5 &    19 &   0 &   1 & SA                      \\
NGC   289       & SB & SB & SB &     0.212 &     0.089 &    11 &    49 &   2 &   1 & SA$\underline{\rm B}$   \\
NGC   428       & SAB & SAB & SB &     0.254 &     0.100 &    19 &    50 &   3 &   1 & SB                      \\
NGC   488       & SA & SA & SA &     0.028 &     0.020 &    11 &    40 &   0 &   0 & SA                      \\
NGC   578       & SAB & SAB & SB &     0.180 &     0.168 &     9 &    37 &   2 &   2 & SAB                     \\
NGC   613       & SB & SB & SB &     0.298 &     0.319 &    39 &    65 &   3 &   3 & SB                      \\
NGC   685       & SAB & SB & SB &     0.389 &     0.157 &     9 &    23 &   4 &   2 & SB                      \\
NGC   864       & SAB & SAB & SB &     0.321 &     0.134 &    13 &    19 &   3 &   1 & SB                      \\
NGC  1042       & SAB & SAB & SAB &     0.044 &     0.530 &     5 &    21 &   0 &   5 & SA                      \\
NGC  1058       & SA & SA & SA &     0.129 &     0.097\rlap{:} &    15 &   ... &   1 &   1 & SAB                     \\
NGC  1073       & SB & SB & SB &     0.561 &     0.264 &    15 &    29 &   6 &   3 & SB                      \\
NGC  1084       & SA & SA & SA &     0.038 &     0.197 &     5 &    23 &   0 &   2 & SA                      \\
NGC  1087       & SAB & SB & SB &     0.428 &     0.265 &     5 &    25 &   4 &   3 & SB                      \\
NGC  1187       & SB & SB & SB &     0.117 &     0.183 &    17 &    31 &   1 &   2 & SAB                     \\
NGC  1241       & SB & SAB & SB &     0.181 &     0.153 &    11 &    19 &   2 &   2 & SAB                     \\
NGC  1300       & SB & SB & SB &     0.524 &     0.184 &    57 &   111 &   5 &   2 & SB                      \\
NGC  1302       & SB & SAB & SB &     0.061 &     0.033 &    17 &    89 &   1 &   0 & S$\underline{\rm A}$B   \\
NGC  1309       & SA & SA & SAB &     0.091 &     0.132 &     9 &    15 &   1 &   1 & S$\underline{\rm A}$B   \\
NGC  1317       & SAB & SB & SB &     0.085 &     0.031 &    35 &    83 &   1 &   0 & S$\underline{\rm A}$B   \\
NGC  1371       & SAB & SAB & SAB &     0.049 &     0.109 &     7 &    19 &   0 &   1 & SA                      \\
NGC  1385       & SB & SB & SB &     0.269 &     0.262 &     3 &    31 &   3 &   3 & SB                      \\
NGC  1493       & SB & SAB & SB &     0.319 &     0.159 &     9 &    19 &   3 &   2 & SB                      \\
NGC  1559       & SB & SB & SB &     0.328 &     0.185 &     5 &    45 &   3 &   2 & SB                      \\
NGC  1617       & SB & SA & SAB &     0.034 &     0.078 &     7 &    35 &   0 &   1 & SA                      \\
NGC  1637       & SAB & SA & SB &     0.193 &     0.066 &    11 &    17 &   2 &   1 & SAB                     \\
NGC  1703       & SB & SA & SAB &     0.073 &     0.097 &     9 &    23 &   1 &   1 & S$\underline{\rm A}$B   \\
NGC  1792       & SA & SA & SA &     0.060 &     0.150 &     5 &    31 &   1 &   2 & S$\underline{\rm A}$B   \\
NGC  1832       & SB & SAB & SB &     0.176 &     0.131 &    11 &    39 &   2 &   1 & SAB                     \\
NGC  2090       & SA & SAB & SA &     0.087 &     0.090 &     9 &    17 &   1 &   1 & S$\underline{\rm A}$B   \\
NGC  2139       & SAB & SB & SB &     0.356 &     0.198 &     3 &    21 &   4 &   2 & SB                      \\
NGC  2196       & SA & SA & SA &     0.069 &     0.094 &     7 &   107 &   1 &   1 & S$\underline{\rm A}$B   \\
NGC  2442       & SAB & SB & SB &     0.412 &     0.600\rlap{:} &    45 &    71 &   4 &   6 & SB                      \\
NGC  2559       & SB & SAB & SB &     0.334 &     0.169 &    25 &    43 &   3 &   2 & SB                      \\
NGC  2566       & SB & SB & SB &     0.270 &     0.220 &    45 &    79 &   3 &   2 & SB                      \\
NGC  2775       & SA & SA & SA &     0.037 &     0.043 &    11 &    33 &   0 &   0 & SA                      \\
NGC  2964       & SAB & SA & SAB &     0.270 &     0.110 &    13 &    24 &   3 &   1 & SB                      \\
NGC  3059       & SB & SB & SB &     0.533 &     0.305 &     8 &   113\rlap{:} &   5 &   3 & SB                      \\
NGC  3166       & SAB & SA & SB &     0.108 &     0.073 &    21 &    77 &   1 &   1 & SAB                     \\
NGC  3169       & SA & SA & SA &     0.089 &     0.036 &    11 &   113\rlap{:} &   1 &   0 & S$\underline{\rm A}$B   \\
NGC  3223       & SA & SA & SA &     0.025 &     0.047 &     7 &    81 &   0 &   0 & SA                      \\
NGC  3227       & SAB & SAB & SB &     0.151 &     0.078 &    21 &    43 &   2 &   1 & SAB                     \\
NGC  3261       & SB & SAB & SB &     0.166 &     0.100 &    15 &    30 &   2 &   1 & SAB                     \\
NGC  3275       & SB & SB & SB &     0.183 &     0.166 &    19 &   102 &   2 &   2 & SAB                     \\
NGC  3319       & SB & SB & SB &     0.537 &     0.309 &     9 &    75 &   5 &   3 & SB                      \\
NGC  3338       & SA & SAB & SAB &     0.049 &     0.076 &     5 &    33 &   0 &   1 & SA                      \\
NGC  3423       & SA & SA & SA &     0.037 &     0.163 &     7 &    63 &   0 &   2 & SA                      \\
NGC  3504       & SAB & SB & SB &     0.286 &     0.069 &    19 &    33 &   3 &   1 & SB                      \\
NGC  3507       & SB & SB & SB &     0.188 &     0.098 &    13 &    19 &   2 &   1 & SAB                     \\
NGC  3513       & SB & SB & SB &     0.521 &     0.293 &    13 &    47 &   5 &   3 & SB                      \\
NGC  3583       & SB & SAB & SB &     0.170 &     0.189 &     9 &    17 &   2 &   2 & SAB                     \\
NGC  3593       & SA & SA & SA &     0.151 &     0.010 &     7 &    60 &   2 &   0 & SAB                     \\
NGC  3596       & SAB & SA & SAB &     0.080 &     0.200 &     7 &    30 &   1 &   2 & S$\underline{\rm A}$B   \\
NGC  3646       & SI & SA & SAB &     0.081 &     0.260 &     5 &    42 &   1 &   3 & S$\underline{\rm A}$B   \\
NGC  3675       & SA & SAB & SB &     0.078 &     0.083 &    11 &    49 &   1 &   1 & S$\underline{\rm A}$B   \\
NGC  3681       & SAB & SAB & SB &     0.187 &     0.070 &     5 &    19 &   2 &   1 & SAB                     \\
NGC  3684       & SA & SAB & SAB &     0.086 &     0.163 &     3 &   113 &   1 &   2 & S$\underline{\rm A}$B   \\
NGC  3686       & SB & SAB & SB &     0.225 &     0.082 &     7 &    15 &   2 &   1 & SA$\underline{\rm B}$   \\
NGC  3726       & SAB & SAB & SB &     0.212 &     0.174 &    17 &    41 &   2 &   2 & SA$\underline{\rm B}$   \\
NGC  3810       & SA & SA & SAB &     0.049 &     0.110 &     7 &    11 &   0 &   1 & SA                      \\
NGC  3887       & SB & SAB & SB &     0.093 &     0.175 &     9 &    23 &   1 &   2 & S$\underline{\rm A}$B   \\
NGC  3893       & SAB & SA & SAB &     0.122 &     0.132 &     9 &    47 &   1 &   1 & SAB                     \\
NGC  3938       & SA & SA & SA &     0.022 &     0.052 &    11 &    37 &   0 &   1 & SA                      \\
NGC  3949       & SA & SAB & SAB &     0.171 &     0.269 &     3 &    17 &   2 &   3 & SAB                     \\
NGC  4027       & SB & SB & SB &     0.569 &     0.316 &     3 &    19 &   6 &   3 & SB                      \\
NGC  4030       & SA & SA & SA &     0.020 &     0.059 &     5 &    53 &   0 &   1 & SA                      \\
NGC  4051       & SAB & SB & SB &     0.097 &     0.257 &    23 &    45 &   1 &   3 & S$\underline{\rm A}$B   \\
NGC  4123       & SB & SB & SB &     0.331 &     0.195 &    21 &    31 &   3 &   2 & SB                      \\
NGC  4136       & SAB & SAB & SB &     0.150 &     0.114 &     7 &    17 &   2 &   1 & SAB                     \\
NGC  4138       & SA & S  & S  &     0.039 &     0.035 &     5 &    17 &   0 &   0 & SA                      \\
NGC  4145       & SAB & SAB & SB &     0.427 &     0.124 &     3 &    25 &   4 &   1 & SB                      \\
NGC  4151       & SAB & SB & SB &     0.114 &     0.039 &    43 &    87 &   1 &   0 & SAB                     \\
NGC  4212       & SA & SA & SAB &     0.060 &     0.210 &     5 &    19 &   1 &   2 & S$\underline{\rm A}$B   \\
NGC  4242       & SAB & SB & SB &     0.225 &     0.050 &    29 &    60 &   2 &   1 & SA$\underline{\rm B}$   \\
NGC  4254       & SA & SA & SAB &     0.098 &     0.101 &     9 &    51 &   1 &   1 & S$\underline{\rm A}$B   \\
NGC  4303       & SAB & SB & SB &     0.075 &     0.243 &    13 &    27 &   1 &   2 & S$\underline{\rm A}$B   \\
NGC  4314       & SB & SB & SB &     0.439 &     0.084 &    35 &    61 &   4 &   1 & SB                      \\
NGC  4394       & SB & SB & SB &     0.259 &     0.070 &    21 &    41 &   3 &   1 & SB                      \\
NGC  4414       & SA & SA & SA &     0.088 &     0.143 &     7 &    21 &   1 &   1 & S$\underline{\rm A}$B   \\
NGC  4450       & SA & SA & SB &     0.116 &     0.085 &    25 &    63 &   1 &   1 & SAB                     \\
NGC  4457       & SAB & SA & SB &     0.078 &     0.050 &    19 &    41 &   1 &   1 & S$\underline{\rm A}$B   \\
NGC  4487       & SAB & SAB & SB &     0.178 &     0.070 &     7 &    34 &   2 &   1 & SAB                     \\
NGC  4504       & SA & SA & SB &     0.075 &     0.138 &     7 &    23 &   1 &   1 & S$\underline{\rm A}$B   \\
NGC  4548       & SB & SB & SB &     0.285 &     0.155 &    33 &    51 &   3 &   2 & SB                      \\
NGC  4571       & SA & SA & SA &     0.022 &     0.080 &     3 &    30 &   0 &   1 & SA                      \\
NGC  4579       & SAB & SB & SB &     0.188 &     0.050 &    21 &    49 &   2 &   1 & SAB                     \\
NGC  4580       & SAB & SA & SA &     0.077 &     0.088 &     7 &    13 &   1 &   1 & S$\underline{\rm A}$B   \\
NGC  4593       & SB & SB & SB &     0.263 &     0.104 &    37 &    53 &   3 &   1 & SB                      \\
NGC  4618       & SB & SB & SB &     0.354 &     0.197 &     7 &    67 &   4 &   2 & SB                      \\
NGC  4643       & SB & SB & SB &     0.245 &     0.039 &    27 &    45 &   2 &   0 & SA$\underline{\rm B}$   \\
NGC  4647       & SAB & SB & SB &     0.108 &     0.112 &     7 &    57 &   1 &   1 & SAB                     \\
NGC  4651       & SA & SA & SAB &     0.061 &     0.095 &     7 &    13 &   1 &   1 & S$\underline{\rm A}$B   \\
NGC  4654       & SAB & SAB & SB &     0.136 &     0.144 &     5 &    45 &   1 &   1 & SAB                     \\
NGC  4665       & SB & SB & SB &     0.257 &     0.037 &    25 &    73 &   3 &   0 & SB                      \\
NGC  4689       & SA & SA & SA &     0.050 &     0.067 &    13 &    39 &   1 &   1 & S$\underline{\rm A}$B   \\
NGC  4691       & SB & SB & SB &     0.499 &     0.063 &     9 &    87 &   5 &   1 & SB                      \\
NGC  4698       & SA & SA & SA &     0.088 &     0.059 &    45 &   105 &   1 &   1 & S$\underline{\rm A}$B   \\
NGC  4699       & SAB & SB & SB &     0.138 &     0.030 &     9 &    19 &   1 &   0 & SAB                     \\
NGC  4772       & SA & SA & SB &     0.042 &     0.030 &    45 &    63 &   0 &   0 & SA                      \\
NGC  4775       & SA & SA & SA &     0.105 &     0.125 &     3 &    27 &   1 &   1 & SAB                     \\
NGC  4781       & SB & SAB & SB &     0.205 &     0.312 &     7 &    17 &   2 &   3 & SA$\underline{\rm B}$   \\
NGC  4900       & SB & SB & SB &     0.372 &     0.167 &     5 &    19 &   4 &   2 & SB                      \\
NGC  4902       & SB & SB & SB &     0.272 &     0.060 &    15 &    67 &   3 &   1 & SB                      \\
NGC  4930       & SB & SB & SB &     0.210 &     0.110 &    31 &   109 &   2 &   1 & SA$\underline{\rm B}$   \\
NGC  4939       & SA & SAB & SAB &     0.119 &     0.084 &    11 &    97 &   1 &   1 & SAB                     \\
NGC  4995       & SAB & SAB & SB &     0.203 &     0.207 &    11 &    19 &   2 &   2 & SA$\underline{\rm B}$   \\
NGC  5054       & SA & SA & SAB &     0.065 &     0.088 &    13 &    69 &   1 &   1 & S$\underline{\rm A}$B   \\
NGC  5085       & SA & SA & SAB &     0.155 &     0.109 &    19 &    43 &   2 &   1 & SAB                     \\
NGC  5101       & SB & SB & SB &     0.186 &     0.033 &    39 &   109 &   2 &   0 & SAB                     \\
NGC  5121       & SA & SA & SA &     0.024 &     0.030 &    25 &    57 &   0 &   0 & SA                      \\
NGC  5248       & SAB & SA & SA &     0.061 &     0.270 &     7 &    51 &   1 &   3 & S$\underline{\rm A}$B   \\
NGC  5247       & SA & SA & SA &     0.020 &     0.327 &     3 &    65 &   0 &   3 & SA                      \\
NGC  5334       & SB & SB & SB &     0.322 &     0.145 &     5 &    11 &   3 &   1 & SB                      \\
NGC  5427       & SA & SA & SA &     0.083 &     0.235 &     7 &    33 &   1 &   2 & S$\underline{\rm A}$B   \\
NGC  5483       & SA & SAB & SB &     0.174 &     0.109 &     7 &    19 &   2 &   1 & SAB                     \\
NGC  5643       & SAB & SAB & SB &     0.321 &     0.236 &    27 &    45 &   3 &   2 & SB                      \\
NGC  5676       & SA & SA & SAB &     0.087 &     0.080 &    11 &    23 &   1 &   1 & S$\underline{\rm A}$B   \\
NGC  5701       & SB & SB & SB &     0.139 &     0.053 &    27 &   105 &   1 &   1 & SAB                     \\
NGC  5713       & SAB & SAB & SB &     0.335 &     0.111 &     7 &    15 &   3 &   1 & SB                      \\
NGC  5850       & SB & SB & SB &     0.311 &     0.053 &    39 &    65 &   3 &   1 & SB                      \\
NGC  5921       & SB & SB & SB &     0.255 &     0.349 &    21 &    37 &   3 &   3 & SB                      \\
NGC  5962       & SA & SAB & SB &     0.141 &     0.055 &     9 &    15 &   1 &   1 & SAB                     \\
NGC  6215       & SA & SA & SAB &     0.079 &     0.230 &     3 &    24 &   1 &   2 & S$\underline{\rm A}$B   \\
NGC  6221       & SB & SAB & SB &     0.430 &     0.207 &    25 &    43 &   4 &   2 & SB                      \\
NGC  6300       & SB & SAB & SB &     0.222 &     0.175 &    29 &    63 &   2 &   2 & SA$\underline{\rm B}$   \\
NGC  6384       & SAB & SB & SB &     0.135 &     0.050 &    11 &    35 &   1 &   1 & SAB                     \\
NGC  6753       & SA & SA & SA &     0.029 &     0.032 &     5 &    15 &   0 &   0 & SA                      \\
NGC  6782       & SAB & SAB & SB &     0.163 &     0.030 &    21 &    44 &   2 &   0 & SAB                     \\
NGC  6902       & SA & SA & SB &     0.034 &     0.080 &    11 &    30 &   0 &   1 & SA                      \\
NGC  6907       & SB & SB & SB &     0.071 &     0.329 &     3 &    25 &   1 &   3 & S$\underline{\rm A}$B   \\
NGC  7083       & SA & SA & SA &     0.033 &     0.071 &     5 &    23 &   0 &   1 & SA                      \\
NGC  7217       & SA & SA & SA &     0.033 &     0.036 &     9 &   109 &   0 &   0 & SA                      \\
NGC  7205       & SA & SA & SAB &     0.048 &     0.061 &     7 &    55 &   0 &   1 & SA                      \\
NGC  7213       & SA & SA & SA &     0.004 &     0.024 &    11 &    93 &   0 &   0 & SA                      \\
NGC  7412       & SB & SAB & SAB &     0.060 &     0.434 &    11 &    45 &   1 &   4 & S$\underline{\rm A}$B   \\
NGC  7418       & SAB & SAB & SB &     0.158 &     0.153 &    11 &    35 &   2 &   2 & SAB                     \\
NGC  7479       & SB & SB & SB &     0.702 &     0.260 &    27 &    41 &   7 &   3 & SB                      \\
NGC  7552       & SB & SB & SB &     0.393 &     0.055 &    39 &    65 &   4 &   1 & SB                      \\
NGC  7713       & SB & SA & SA &     0.040 &     0.097 &     5 &    23 &   0 &   1 & SA                      \\
NGC  7723       & SB & SB & SB &     0.319 &     0.120 &    11 &    22 &   3 &   1 & SB                      \\
NGC  7727       & SAB & SAB & SA &     0.087 &     0.145 &     7 &    99 &   1 &   1 & S$\underline{\rm A}$B   \\
NGC  7741       & SB & SB & SB &     0.736 &     0.324 &    11 &    27 &   7 &   3 & SB                      \\
IC   4444       & SAB & SA & SB &     0.254 &     0.140 &     5 &    16 &   3 &   1 & SB                      \\
IC   5325       & SAB & SA & SAB &     0.030 &     0.213 &     5 &    11 &   0 &   2 & SA                      \\
ESO  138- 10    & SA & SA & SA &     0.038 &     0.134 &     7 &    67 &   0 &   1 & SA                      \\
\enddata
\end{deluxetable}

\clearpage

\begin{deluxetable}{lccccccr}
\tabletypesize{\scriptsize}
\tablewidth{0pc}
\tablecaption{Mean Bar and Spiral Strength by Family and Variety}
\tablehead{
\colhead{RC3} &
\colhead{$<Q_b>$} &
\colhead{standard} &
\colhead{mean} &
\colhead{$<Q_s>$} &
\colhead{standard} &
\colhead{mean} &
\colhead{$n$} \\
\colhead{classification} &
\colhead{} &
\colhead{deviation} &
\colhead{error} &
\colhead{} &
\colhead{deviation} &
\colhead{error} &
\colhead{} \\
} 
\startdata
\multicolumn{8}{c}{Full sample}\\
SA  &   0.069  &   0.043  &   0.006 &  0.101 &    0.069  &   0.010   &     48\\
SAB &   0.177  &   0.114  &   0.017 &  0.147 &    0.118  &   0.018   &     45\\
SB  &   0.294  &   0.166  &   0.023 &  0.171 &    0.100  &   0.014   &     53\\
\multicolumn{8}{c}{$T$ $\leq$ 4}\\
SA  &   0.062  &   0.041  &   0.008 &  0.082 &    0.072  &   0.014   &     28\\
SAB &   0.143  &   0.100  &   0.020 &  0.134 &    0.127  &   0.025   &     26\\
SB  &   0.247  &   0.131  &   0.022 &  0.143 &    0.101  &   0.017   &     36\\
\multicolumn{8}{c}{$T$ $>$ 4}\\
SA  &   0.080  &   0.044  &   0.010 &  0.127 &    0.054  &   0.012   &     20\\
SAB &   0.225  &   0.117  &   0.027 &  0.165 &    0.105  &   0.024   &     19\\
SB  &   0.395  &   0.191  &   0.046 &  0.228 &    0.069  &   0.017   &     17\\
\multicolumn{8}{c}{SA Galaxies}\\
r   &   0.047  &   0.037  &   0.012 &  0.066 &    0.058  &   0.019   &      9\\
rs  &   0.077  &   0.025  &   0.008 &  0.111 &    0.034  &   0.011   &      9\\
s   &   0.074  &   0.047  &   0.009 &  0.109 &    0.077  &   0.014   &     30\\
\multicolumn{8}{c}{SAB Galaxies}\\
r   &  0.199   &   0.094  &   0.033 &  0.092 &    0.056  &   0.020   &      8\\
rs  &  0.167   &   0.119  &   0.022 &  0.164 &    0.105  &   0.020   &     29\\
s   &  0.193   &   0.124  &   0.044 &  0.141 &    0.189  &   0.067   &      8\\
\multicolumn{8}{c}{SB Galaxies}\\
r   &  0.201   &   0.107  &   0.029 &  0.135 &    0.081  &   0.022   &     14\\
rs  &  0.327   &   0.139  &   0.029 &  0.177 &    0.093  &   0.019   &     23\\
s   &  0.329   &   0.215  &   0.054 &  0.193 &    0.119  &   0.030   &     16\\
\enddata
\end{deluxetable}

\clearpage

\begin{deluxetable}{lc}
\tabletypesize{\scriptsize}
\tablewidth{0pc}
\tablecaption{Definitions of ``$Q_b$ Families"}
\tablehead{
\colhead{Family} &
\colhead{Range}
} 
\startdata
SA & $Q_b$ $<$ 0.05 \\
S$\underline{\rm A}$B &  0.05 $\leq Q_b <$ 0.10 \\
SAB &  0.10 $\leq Q_b <$ 0.20 \\
SA$\underline{\rm B}$ &  0.20 $\leq Q_b <$ 0.25 \\
SB & $Q_b$ $\geq$ 0.25 \\
\enddata
\end{deluxetable}

\clearpage

\begin{deluxetable}{lccl}
\tabletypesize{\scriptsize}
\tablewidth{0pc}
\tablecaption{SA Galaxies Classified as SB in Near-IR by Eskridge et al. (2002)}
\tablehead{
\colhead{Name} &
\colhead{RC3 Family} &
\colhead{OSU $H$ Family} &
\colhead{$Q_b$ Family} 
} 
\startdata
NGC 3675 & SA & SB & S$\underline{\rm A}$B \\
NGC 4450 & SA & SB & SAB \\
NGC 4504 & SA & SB & S$\underline{\rm A}$B \\
NGC 5483 & SA & SB & SAB \\
NGC 5962 & SA & SB & SAB \\
NGC 6902 & SA & SB & SA \\
\enddata
\end{deluxetable}

\begin{deluxetable}{lrrrrr}
\tabletypesize{\scriptsize}
\tablewidth{0pc}
\tablecaption{General Comparison of $Q_b$ Family with Other Bar Classifications}
\tablehead{
\colhead{} &
\colhead{SA} &
\colhead{S$\underline{\rm A}$B} &
\colhead{SAB} &
\colhead{SA$\underline{\rm B}$} &
\colhead{SB} 
} 
\startdata
\multicolumn{6}{c}{}\\
RC3 SA      &  21  &   18  &    9   &   0   &    0\\
RC3 SAB     &   5  &    9  &   16   &   3   &   12\\
RC3 SB      &   2  &    6  &    9   &   6   &   31\\
\multicolumn{6}{c}{}\\
OSU $B$ SA  &   23  &   23  &    8  &   0   &    2\\
OSU $B$ SAB &    3  &    6  &   16  &   5   &    9\\
OSU $B$ SB  &    1  &    4  &   10  &   4   &   32\\
\multicolumn{6}{c}{}\\
OSU $H$ SA  &   20  &   12  &    3  &    0  &    1\\
OSU $H$ SAB &    5  &   12  &    4  &    0  &    1\\
OSU $H$ SB  &    2  &    9  &   27  &    9  &   41\\
\multicolumn{6}{c}{}\\
without Fourier bar  &    26  &   23  &    6  &    1  &    2\\
with Fourier bar     &     2  &   10  &   28  &    8  &   41\\
\enddata
\end{deluxetable}

\clearpage

\begin{figure}
\figurenum{2 (cont.)}
\plotone{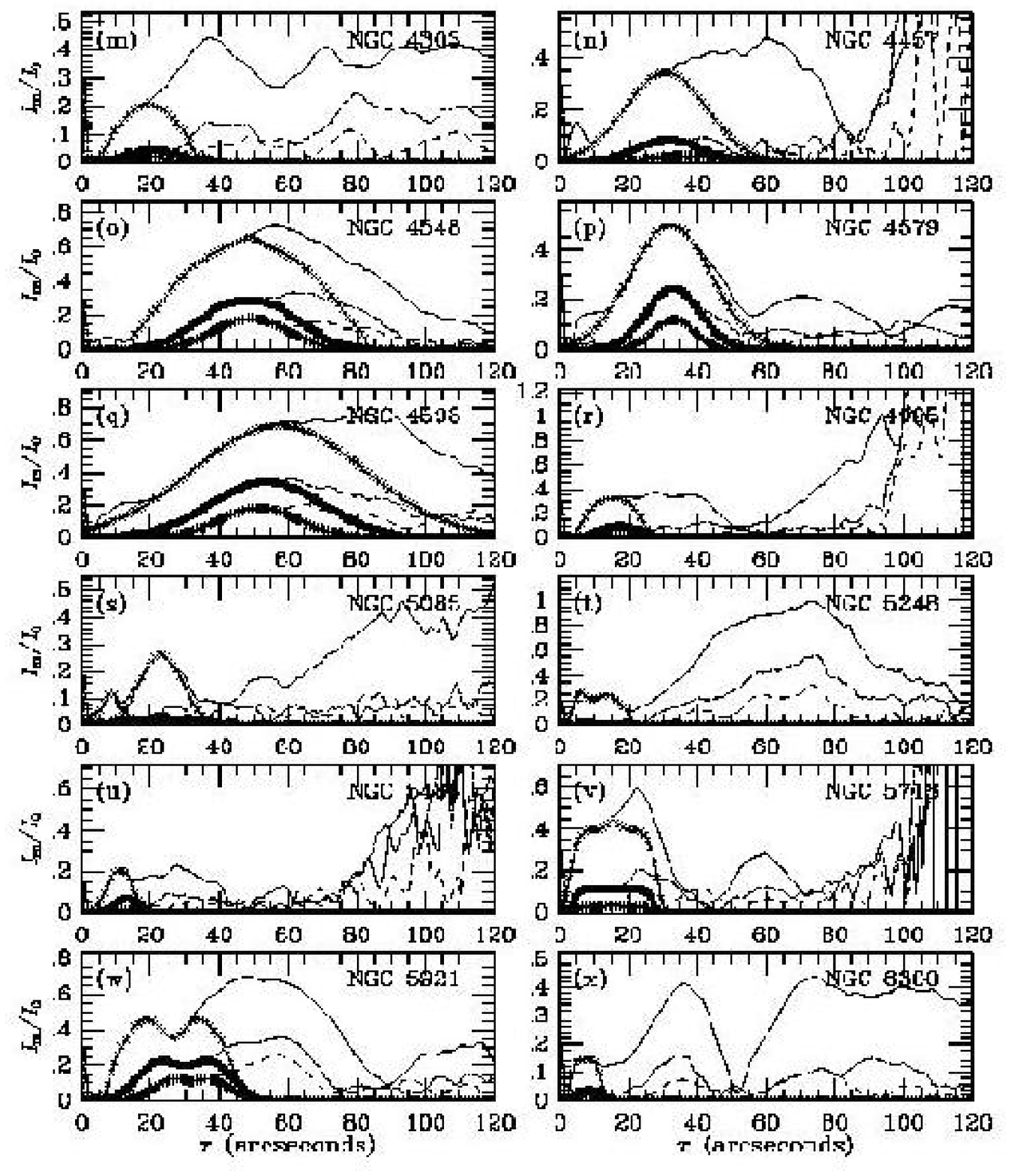}
\caption{}
\end{figure}
\begin{figure}
\figurenum{2 (cont.)}
\caption{Example plots of relative Fourier intensity amplitudes as a function of
radius for 24 OSUBGS galaxies. Symbols show the 
extrapolations used for our analysis (see text). 
For each case, even terms for $m$=2 (solid curve), 4
(dotted curve), and 6 (short dashed curve)
are shown.}
\end{figure}

\clearpage

\begin{figure}
\figurenum{3}
\plotone{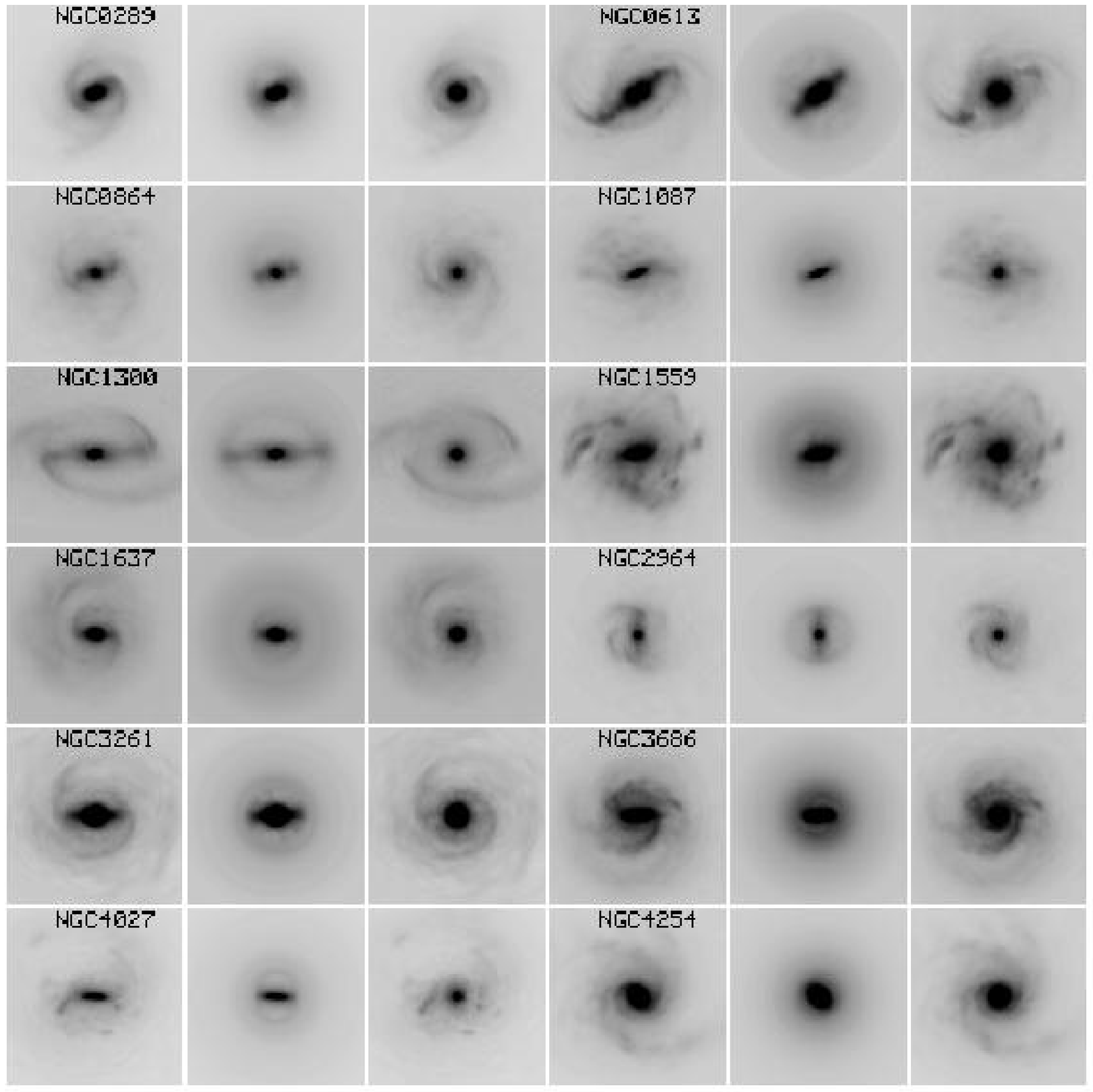}
\caption{}
\label{separations}
\end{figure}
\begin{figure}
\figurenum{3 (cont.)}
\plotone{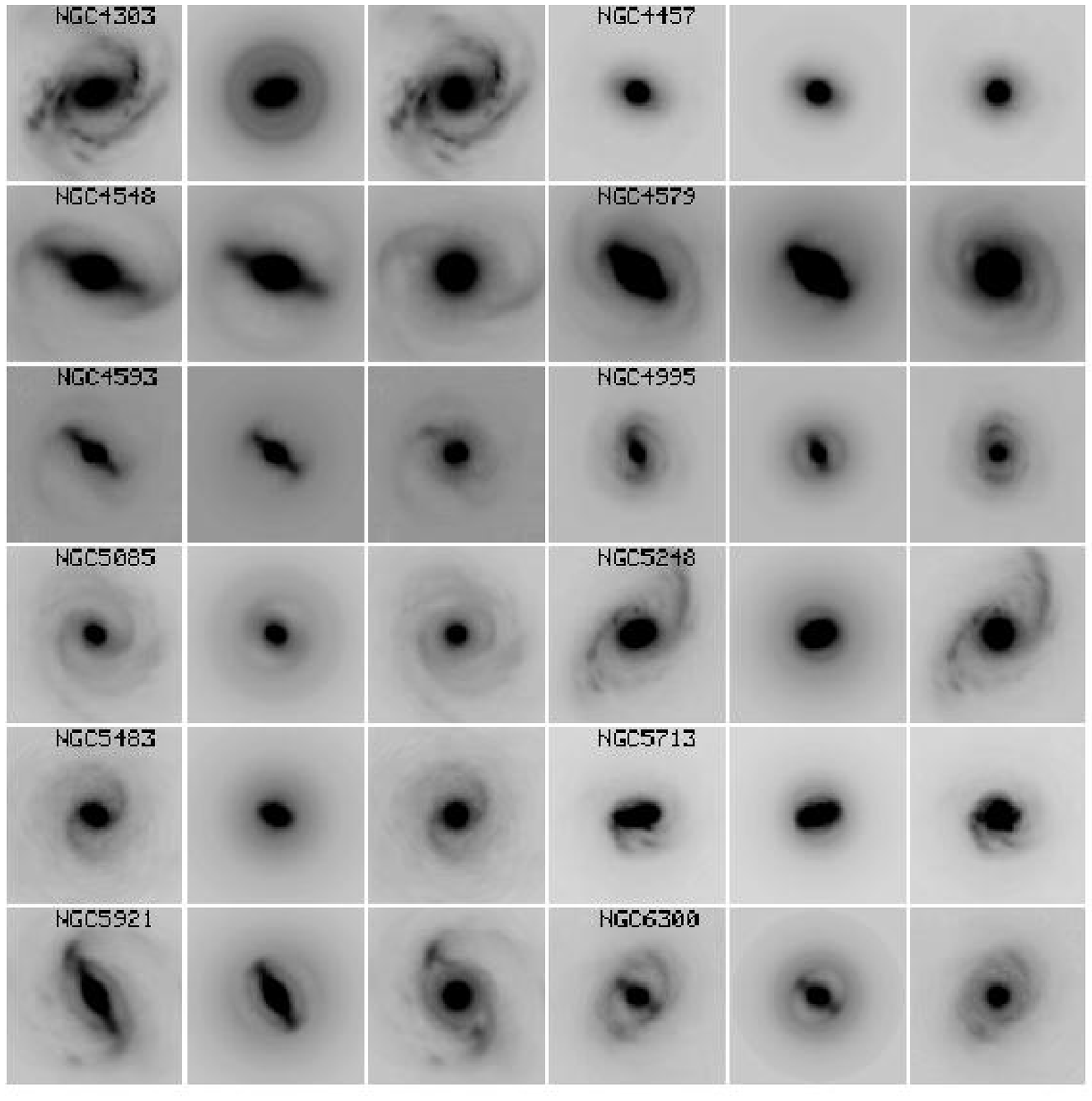}
\caption{}
\end{figure}
\begin{figure}
\figurenum{3 (cont.)}
\caption{Illustration of the bar/spiral separations for the same 24 galaxies
as in Figure 1, using the extrapolations shown in that figure. Three images
are shown for each galaxy: the total $m$=0-20 Fourier-smoothed image (left),
the bar plus disk image (middle), and the spiral plus disk image (right).}
\end{figure}

\clearpage

\begin{figure}
\figurenum{4}
\caption{Plots of maximum relative torques $Q_T(r)$ versus radius $r$ for the bar 
and spiral of NGC 6951, from BBK03, illustrating the definitions of $Q_b$,
$Q_s$, and $Q_g$ and $r(Q_b)$, $r(Q_s)$, and $r(Q_g)$.}
\plotone{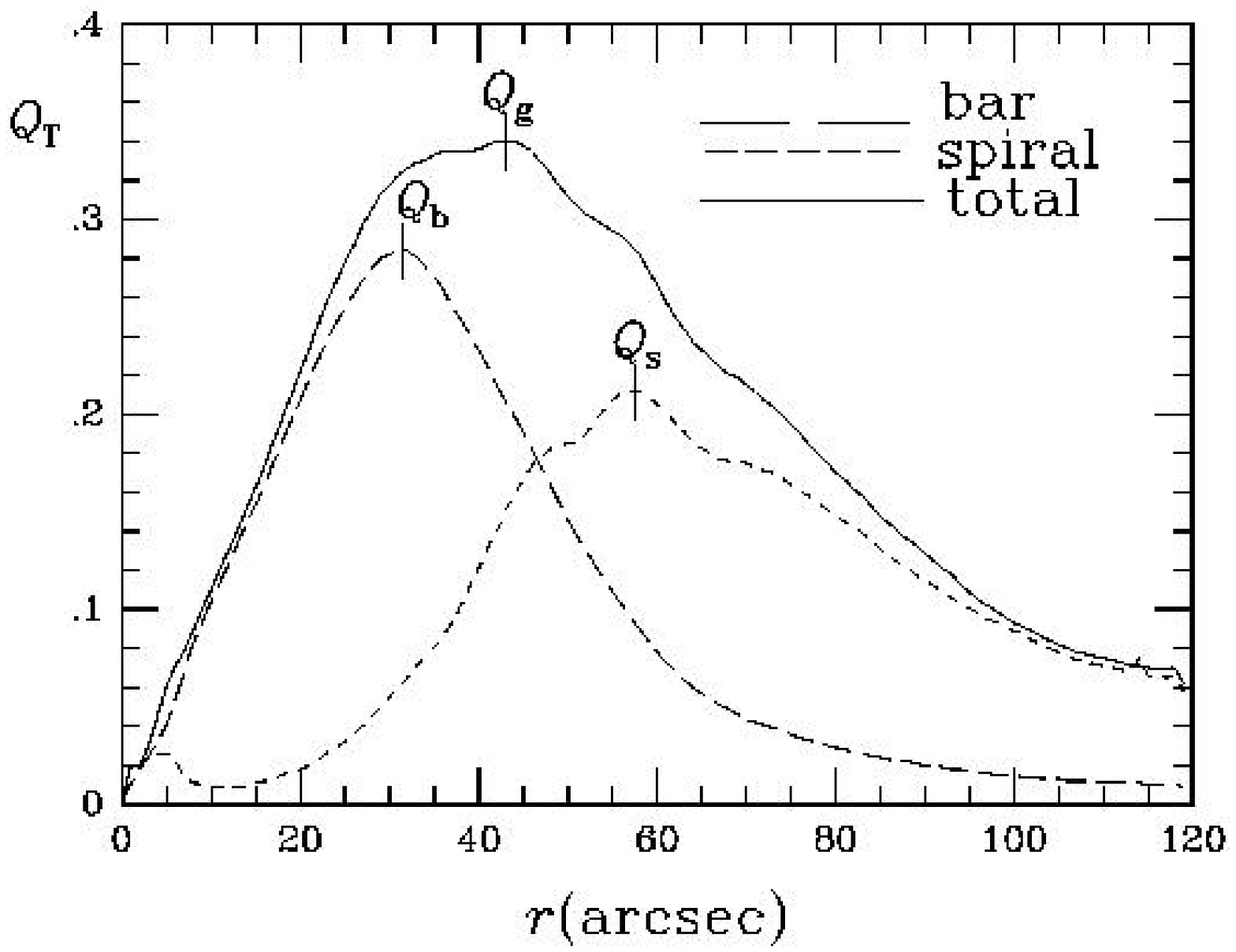}
\label{schematic}
\end{figure}

\begin{figure}
\figurenum{5}
\plotone{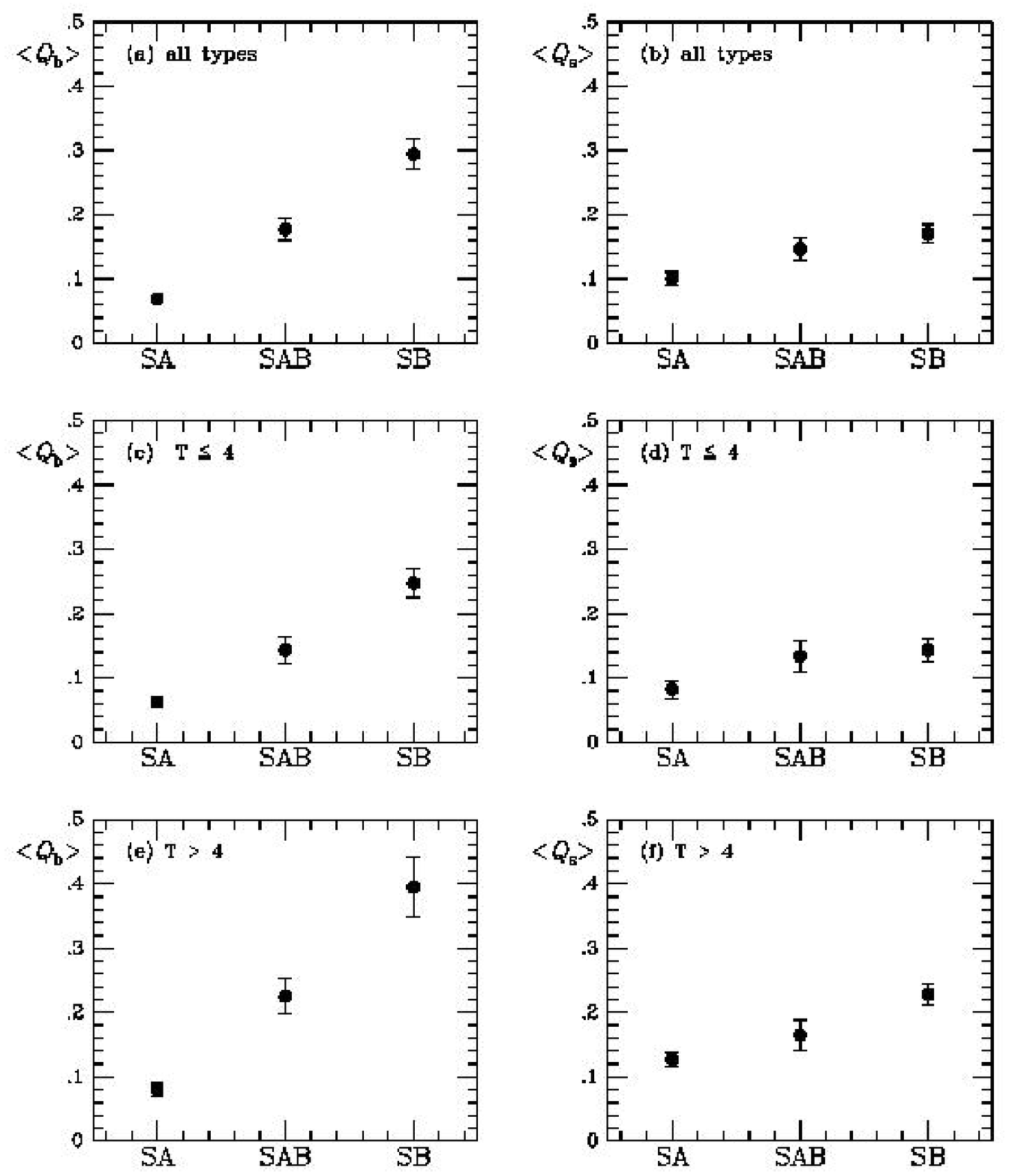}
\caption{Plots of $<Q_b>$ and $<Q_s>$ for 146 OSUBGS galaxies (excluding NGC 3646,
classified as a ring galaxy in RC3):
(a,b) over all spiral types; (c,d) for types at or earlier than Sbc ($T$=4); 
and (e,f) for types later than Sbc. The data illustrated are compiled in Table 2.
The error bars are mean errors.}
\label{byfam}
\end{figure}

\begin{figure}
\figurenum{6}
\plotone{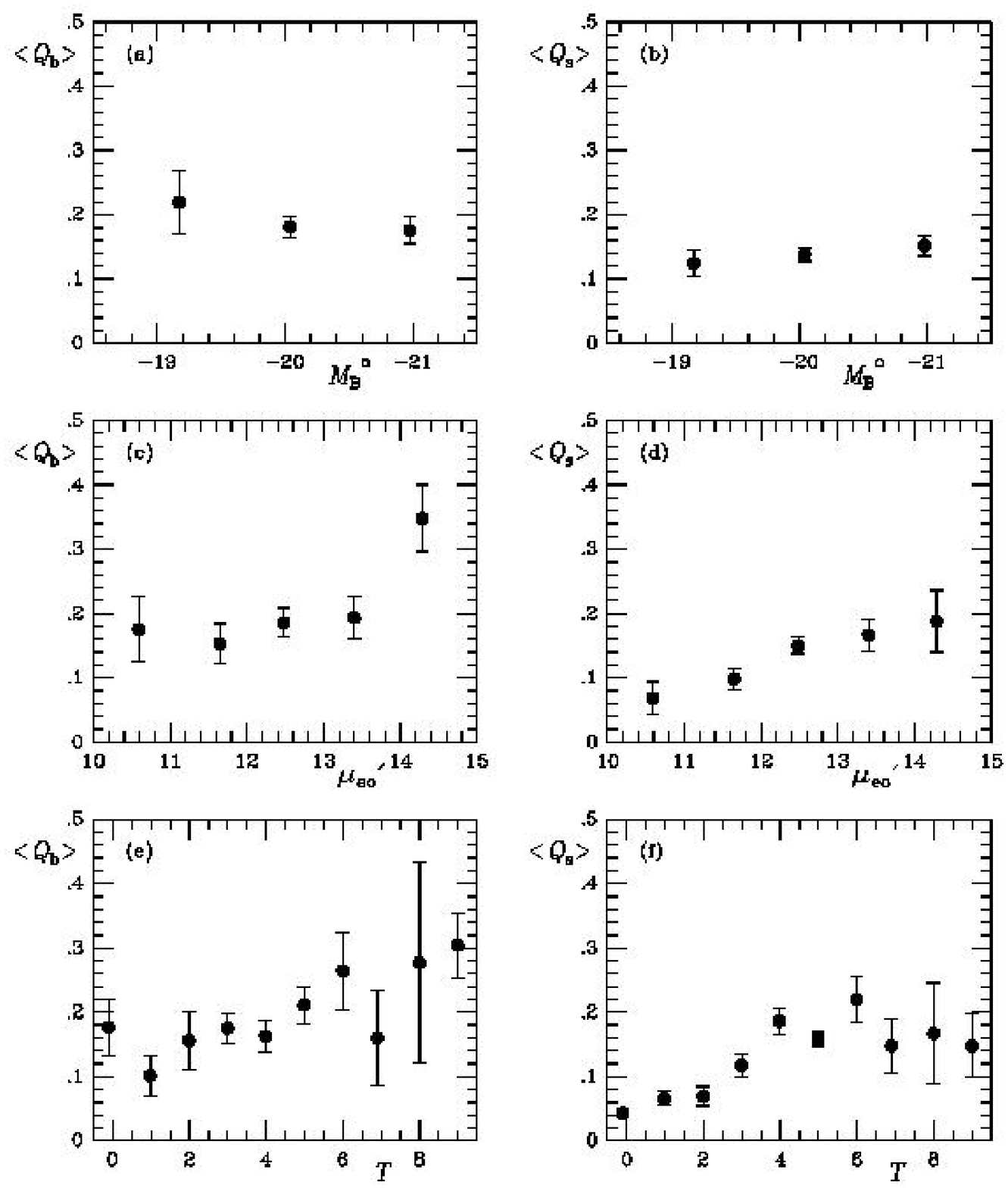}
\caption{Plots of $<Q_b>$ and $<Q_s>$ 
as a function of: (a,b) absolute blue total magnitude $M_B^o$ ($n$=147 galaxies); 
(c,d) photoelectrically determined mean effective surface
brightness in RC3, corrected for tilt and Galactic extinction ($n$=113 galaxies);
and (e,f) RC3 revised Hubble type index ($n$=147 galaxies). The error bars are mean errors.}
\label{byparams}
\end{figure}

\begin{figure}
\figurenum{7}
\plotone{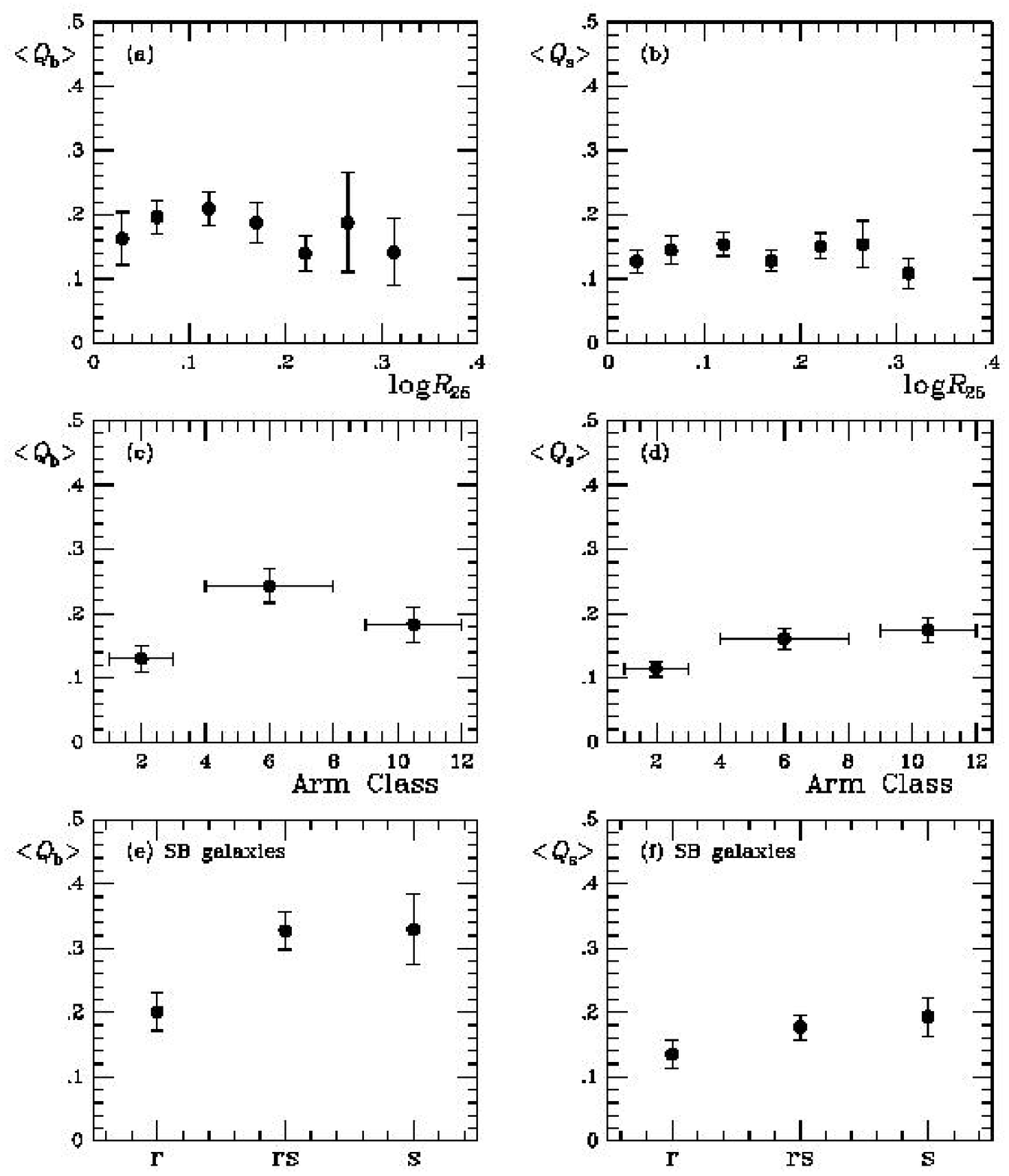}
\caption{Plots of $<Q_b>$ and $<Q_s>$ as a function of: (a,b) RC3 logarithmic
isophotal axis ratio at the $\mu_B$=25.0 mag arcsec$^{-2}$ surface brightness
level ($n$=144 galaxies); (c,d) spiral Arm Class (Elmegreen \& Elmegreen
1987; $n$=107 galaxies); and (e,f) SB spiral variety ($n$=53 galaxies).}
\label{bymoreparams}
\end{figure}

\begin{figure}
\figurenum{8}
\plotone{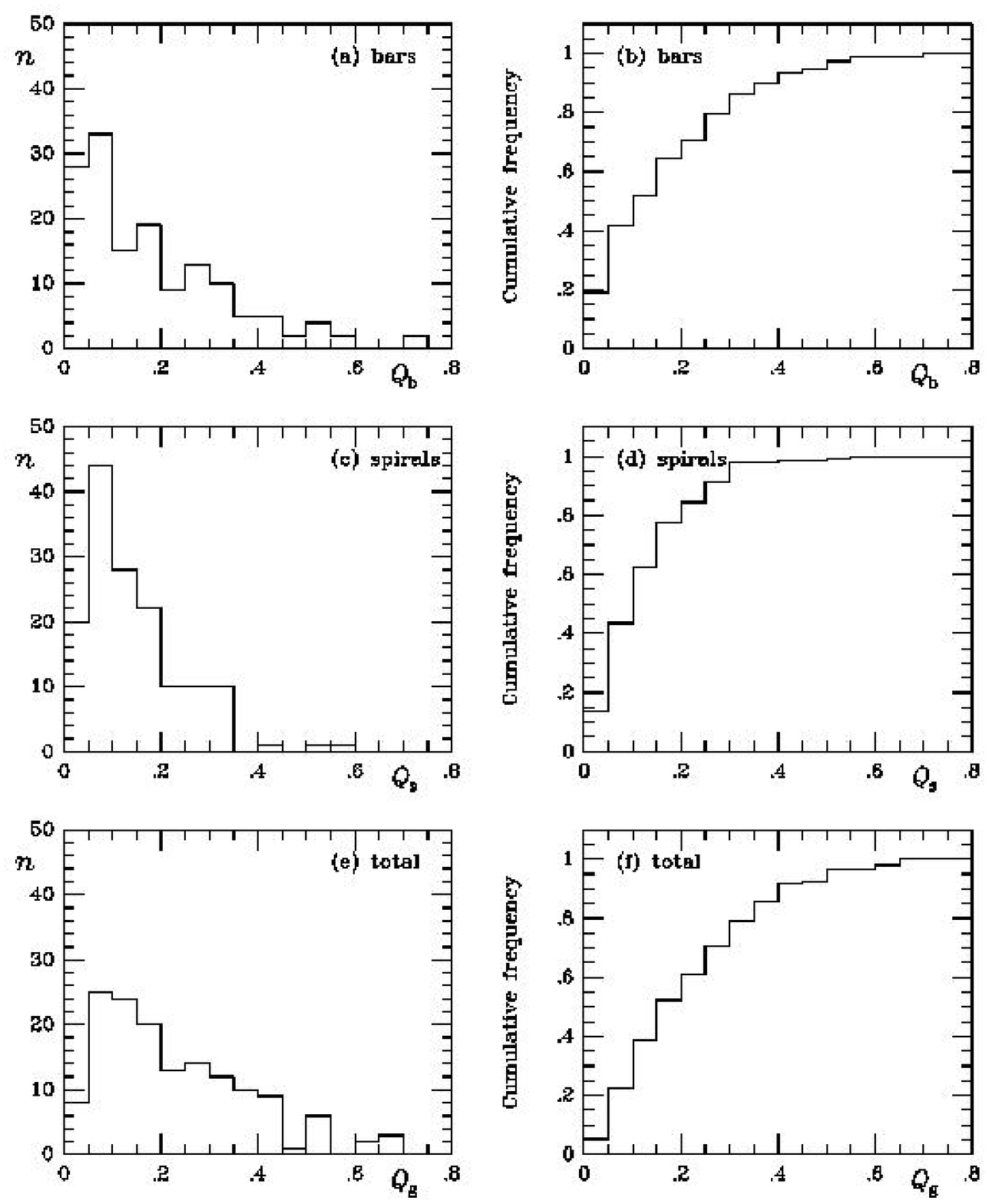}
\caption{Histograms of the distributions of (a,b) bar strength $Q_b$,
(c,d) spiral strength $Q_s$, and (e,f) total nonaxisymmetric strength
$Q_g$, for 147 OSUBGS galaxies less inclined than 65$^{\circ}$.
The cumulative histograms are normalized to the total number of
galaxies. The $Q_g$ data are from LSBV04.}
\label{histos}
\end{figure}

\begin{figure}
\figurenum{9}
\plotone{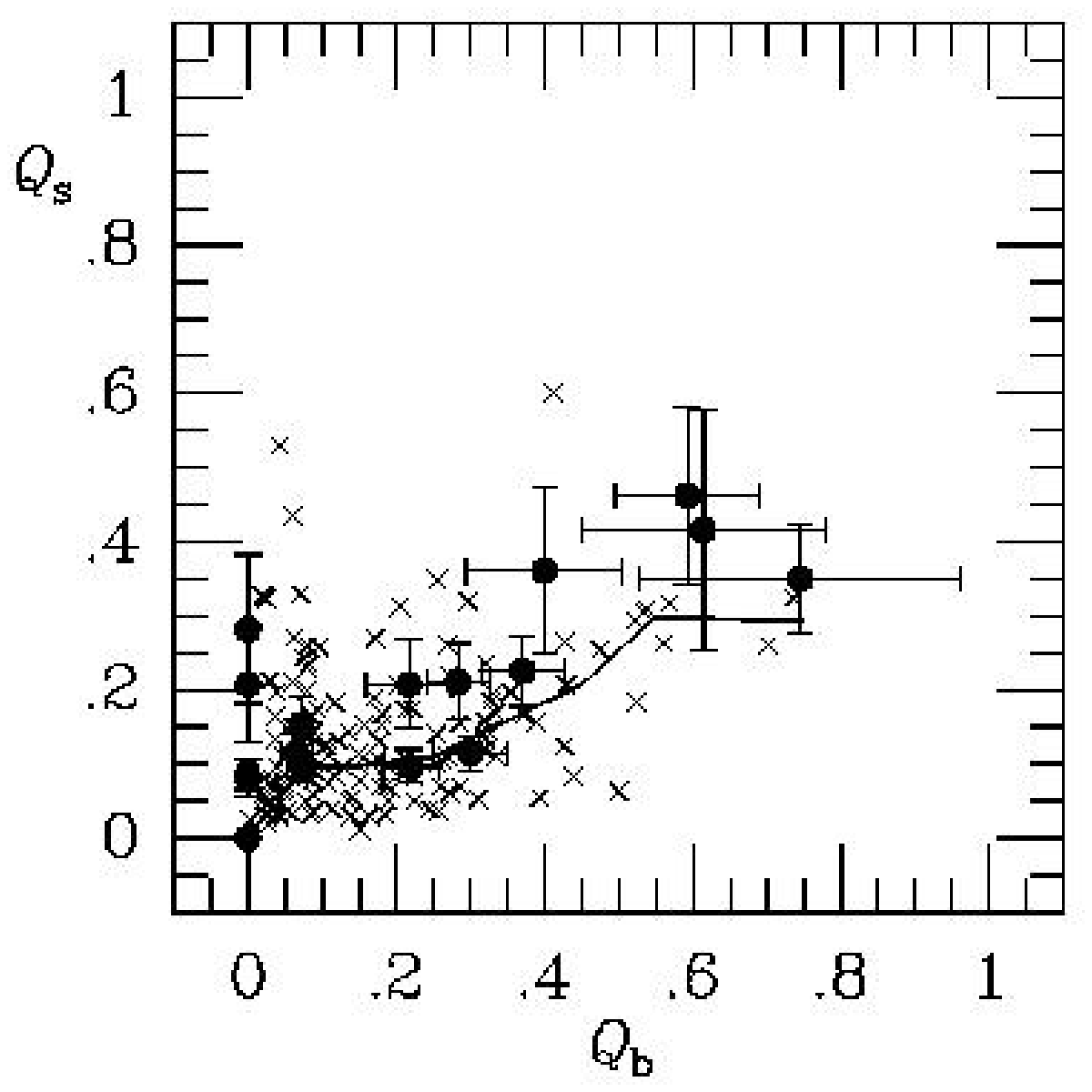}
\caption{Plot of spiral strength $Q_s$ versus bar strength $Q_b$ for
147 OSUBGS galaxies ({\it crosses}) and 17 nearby spirals from
Block et al. (2004; {\it filled circles}). The solid curve shows the
median $Q_s$ for steps of 0.1 in $Q_b$.
}
\label{qsqb}
\end{figure}

\end{document}